\documentclass{jfm}

\usepackage{lineno,hyperref}
\modulolinenumbers[5]

\usepackage{euscript,amsmath,amssymb,float}
\usepackage{rotating,float,subfigure,graphics,threeparttable}

\usepackage{pstricks,rotating,color}
\definecolor{gr}{cmyk}{0.,0.,0,0.25}

\usepackage{natbib}

\newcommand\lae[1]{\label{#1}}

\title{Drag and thermophoresis on a sphere in a rarefied gas based on the
 Cercignani-Lampis model of gas-surface interaction}

\author{Denize Kalempa\aff{1}
  \corresp{\email{kalempa@usp.br}} \and
Felix Sharipov\aff{2}}

\affiliation{\aff{1} Departamento de Ci\^encias B\'asicas e Ambientais, Escola de
Engenharia de Lorena, Universidade de S\~ao Paulo, 12602-810, Lorena, Brazil
\aff{2}Departamento de F\'{\i}sica, Universidade Federal do
Paran\'a, Caixa Postal 19044, 81531-990, Curitiba, Brazil}

\begin{document}
\maketitle

\begin{abstract}

In the present work, the influence of the gas-surface interaction law 
on the classical problems of viscous drag and thermophoresis on a spherical particle
with high thermal conductivity immersed in a monatomic rarefied gas is
investigated on the basis of the solution of a kinetic model to the linearized
Boltzmann equation. The scattering kernel proposed by Cercignani and
Lampis is employed to model the gas-surface interaction law via the setting
of two accommodation coefficients, namely the
tangential momentum accommodation coefficient (TMAC) and the normal energy accommodation
coefficient (NEAC). The viscous drag and thermophoretic forces acting on the
sphere are calculated in a wide range of the rarefaction parameter, which is 
defined as the ratio of the sphere radius to an equivalent 
free path of gaseous particles, so that the free molecular, transition and continuum
regimes of the gas flow are covered. In the free molecular regime the
problem is solved analytically via the method of
the characteristics to solve the collisionless kinetic
equation, while in the transition and continuum regimes the discrete
velocity method is employed to solve the kinetic equation numerically. 
The numerical calculations are carried out in
a range of accommodation coefficients which covers most situations encountered in
practice. The macroscopic characteristics of the gas flow around the
sphere, namely the density and temperature deviations from thermodynamic
equilibrium far from the sphere, the bulk velocity and the heat flux are
calculated and their profiles as functions of the radial distance from the sphere are
presented for some values of rarefaction parameter and accommodation
coefficients. The results show the appearance of the negative thermophoresis
in the near continuum regime and the dependence of this phenomenon on the accommodation
coefficients. To verify the reliability of the calculations, the reciprocity
relation between the cross phenomena which is valid at arbitrary distance from the
sphere was found and then verified numerically within an accuracy of 0.1\%.
The results for the thermophoretic force are compared to the more recent experimental data
found in the literature for a copper sphere in argon gas.  

\end{abstract}

\section{Introduction}

Problems regarding viscous drag and 
thermophoresis on spherical particles immersed in a rarefied gas 
are classical in the field of rarefied
gas dynamics and have been investigated by many
authors over the years, see e.g. \cite{Yam08,Tak07,Loy41,Ber09,Tak08,Che25}.
The study of this topic is motivated by its fundamental importance
for the understanding of the physics underlying some phenomena, such as the
transport of aerosols in the atmosphere, and for practical applications 
such as the development of technologies in the fields of microfluidics, 
semiconductor industry, security of nuclear plants, etc. The well known equations of continuum
mechanics, namely the Navier-Stokes-Fourier equations, see e.g. \cite{Lan05}, can be used to
calculate the drag and the thermophoretic forces acting on a sphere, as well
as the macroscopic characteristics of the gas flow around it, only in situations
where the molecular mean free path is significantly smaller than a
 characteristic length of the gas flow domain so that the continuum hypothesis is
still valid. The Knudsen number ($Kn$), defined as the ratio of the
molecular mean free path to a characteristic lenght of the gas flow, is the parameter
often used to classify the gas flow regimes. The equations of continuum
mechanics are valid when $Kn \ll 1$. For instance, in air at
standard conditions, the molecular mean free path is approximately 0.065 $\mu$m. Then, 
for small particles originated from several sources moving through the air, 
the Knudsen number varies from about zero to 65 when the size of particles
ranges from 100 $\mu$m to 10$^{-3}$ $\mu$m. Therefore, the modelling of the gas flow
around aerosols in the atmosphere, as well as the movement of these particles
itself, cannot be accurately described by the equations of continuum mechanics.
Moreover, even in the continuum regime, the Navier-Stokes-Fourier equations
cannot predict the negative thermophoresis, which
means the movement of aerosol particles from cold to hot regions. This
phenomenon was first predicted theoretically by \cite{Son04} for aerosol particles with high
thermal conductivity related to that of the carrier gas. However, 
experimental data regarding this phenomenon is still scarse in
the literature because its detection is very difficult. Actually, the more
recent data concerning negative thermophoresis is provided by \cite{Bos02},
in which the thermophoretic force on a copper sphere in argon gas was measured in 
a wide range of the gas rarefaction. 

Historically, the viscous drag force on a sphere was first
investigated by \cite{Stokes1845} via hydrodynamic analysis based on the
Navier-Stokes-Fourier equations, with the derivation of his famous formula for the
drag force on a sphere in a slow flow, see e.g. \cite{Lan05}. 
Regarding the thermophoresis, the first attempt to
calculate the thermal force on a sphere in a gas with a temperature gradient
was done by \cite{Epstein1924}. Since the theories of both Stokes and Epstein
were valid in the continuum regime, many attempts to modify the equations of continuum
mechanics as well as the boundary conditions were proposed over the years to 
increase their range of applicability in the Knudsen number. For instance, the correction factor 
proposed by \cite{Cunningham1910} to consider the non-continuum effects of gas slippage
on the boundary was incorporated in the Stokes formula so that its
 applicability was extended to the so called slip flow regime. Concerning
the thermophoresis, continuum
analysis based on the Navier-Stokes-Fourier equations with slip corrections in
the boundary condition was first carried out by \cite{Bro06} in an attempt to improve
the previous theory proposed by Epstein. Methods based on the use of higher-order kinetic theory 
approximations, as that
first proposed by \cite{Gra01}, were also employed to solve the problems of drag and
thermophoresis on a sphere. For instance, \cite{Son04} obtained an
expression for the thermophoretic force acting on a sphere with uniform
temperature corrected up to
the second order in the Knudsen number using an asymptotic theory for small
Knudsen numbers and predicted the negative thermophoresis. 
In a more recent paper, \cite{Torrilhon2010} 
investigated a slow flow past a sphere on the basis of the regularized 13-moment
equations as proposed by \cite{Torrilhon2003}, a method which relies on the combination of
the moment approximation and asymptotic expansion in kinetic theory of gases.
Similarly, \cite{Pad01} investigated the thermophoresis on a sphere by
employing the same method and predicted the negative thermophoresis.
 According to \cite{Torrilhon2010} and \cite{Pad01}, the regularized 13-moment method can be
used to describe the drag and the thermophoresis on a sphere when $Kn < 1$. 
In fact, although many efforts have been done
over the years to expand the validity of the continuum models in the
description of gas flows, it is well known that all the theories and
methods currently available fail in describing gas flows properly when
 $Kn \sim 1$ or $Kn \gg 1$. In these kinds of
situations, corresponding to transition and free molecular regimes, the problem
must be solved at the microscopic level via the methods of rarefied gas
dynamics, which are based on either the solution of the Boltzmann equation,
see e.g. \cite{CerB1, Sha02B}, and its related kinetic models, e.g. \cite{Bha01,Shk02}, 
or the Direct Simulation Monte Carlo method as pioneered by \cite{Bir02}. 

Although an extensive literature concerning the topic under
investigation in the whole range of the Knudsen number based on kinetic
theory is available, most of the papers rely on the assumption of diffuse
reflection or complete accommodation of gas molecules on the 
surface, see e.g. the reviews on thermophoresis by
 \cite{Zhe02} and \cite{You02}. However, in
practice, the assumption of complete accommodation of gas molecules on the surface
 is not always valid and its use can lead to large
deviations of theoretical predictions from experimental data. 
As pointed out by \cite{Zhe02}, actually the gas-surface interaction law is most 
probably something between the widely used diffuse and specular reflection models.
 Thus, the so called accommodation coefficients on the surface 
should be conveniently introduced to accurately describe the gas-surface interaction.
From our knowledge, \cite{Ber09,Ber10} were the first authors to study the
influence of the gas-surface interaction law on the drag and thermophoretic
forces acting on a sphere with basis on a kinetic model to the Boltzmann
equation in the whole range of the Knudsen number. These authors solved
the linearized kinetic equation proposed by \cite{Shk02} numerically via the integral-moment
method, with boundary condition written in terms of accommodation
coefficients of momentum and energy as proposed by \cite{Shenew}, with the distribution
function of reflected molecules from the surface approximated by an
expansion in terms of Hermite polynomials and unknown accommodation coefficients
determined from the conservation laws of momentum and energy on the particle
surface. The qualitative results
presented by the authors show a strong dependence of the drag and 
thermophoretic forces on the accommodation coefficients.    

 Currently, the scattering kernel proposed by \cite{Cer11} is considered the most reliable to model
 the gas-surface interaction law because it provides a correct physical description
of many transport phenomena in gases which are not described correctly by
other models of gas-surface interaction available in the literature. For instance, several experimental
data show us that the exponent which appear in the thermomolecular pressure difference at low
pressures varies from 0.4 to 0.5, see e.g. \cite{Pod01,Edm01}. However, as pointed out by
\cite{Sha45}, the widely used diffuse-specular model proposed by Maxwell for 
the gas-surface interaction cannot explain the reason for the deviation of such exponent from 0.5 in the
free molecular regime. On the contrary, the model proposed by
Cercignani-Lampis provides a more correct description of this phenomenon. 
Concerning the problem of thermophoresis and drag on a sphere, the results presented
recently by \cite{Che25} in the free molecular regime show a qualitative
disagreement between the values of forces obtained from different scattering kernels.
As pointed out by the authors, since all the experimental data available in the literature were obtained 
for small and intermediate Knudsen number, experiments in the free molecular
regime should be carried out to verify which model is closer to reality.
 In the Cercignani-Lampis model of gas-surface interaction, two independent accommodation
coefficients are introduced, namely the tangential momentum accommodation
coefficient (TMAC) and the normal energy accommodation coefficient (NEAC).
This model assumes that the NEAC ranges from 0 to 1, while the TMAC ranges from 0 to 2.
In practice, the values of these accommodation coefficients extracted from
experiments can be found in the literature, see e.g. \cite{Sem01,Tro01,Sha36,
Sha113} for several gases and surfaces. For instance, according to
\cite{Tro01,Sha113}, the NEAC ranges from 0 to 0.1 for Helium and from 0.5 to
0.95 for Argon at ambient temperature and several different smooth mettalic surfaces such as
aluminum, platinum and stainless steel. Moreover, the TMAC of Helium and
Argon ranges from 0.5 to 1 at the same conditions. According to the results
presented by \cite{Che25}, in the free molecular regime the
Cercignani-Lampis model leads the thermophoretic force to increase in case
of incomplete NEAC and to decrease in case of incomplete TMAC. Moreover,
this model leads the drag force to decrease as the accommodation
coefficients decrease. 

In the present work, the influence of the gas-surface interaction law on the
drag and thermophoretic forces acting on a sphere of high thermal
conductivity immersed in a monatomic rarefied gas is investigated by employing the 
Cercignani-Lampis model of
gas-surface interaction. The kinetic equation proposed by \cite{Shk02} 
is solved numerically via a discrete velocity method which
takes into account the discontinuity of the distribution function of
molecular velocities around a convex body, see e.g. \cite{Son02,Son13}. The
viscous drag and the thermophoretic forces on the sphere, as well as the macroscopic 
characteristics of the gas flow around it, are calculated in a wide range of the gas rarefaction so
that the free molecular, transition and hydrodynamic regimes are covered. 
The values of the TMAC and NEAC 
are chosen with basis on experimental data as given by \cite{Tro01,Sha113}.
The reciprocity relation between the cross phenomena is obtained and 
applied as an accuracy criterion for the numerical calculations.
The results obtained for both the drag and thermophoretic forces on the sphere in the whole range of the
gas rarefaction are compared to those results provided by \cite{Ber09,Ber10,Tak07,Tak09}
in case of diffuse scattering on the surface. Moreover, the results obtained for
the forces in the free molecular regime are compared to those presented
by \cite{Che25} in a wide range of TMAC and NEAC. 

Regarding the comparison with experimental data, it is worth mentioning that 
although many data are available in the literature, such a comparison is
still a difficult task because in most of the experiments the carrier gas is air or
a polyatomic gas, and the results are limited to a certain range of 
the Knudsen number which usually covers the continuum and near-continuum
regimes. Moreover, the experiments involve particles of different materials
and some physical properties of matter, such as the thermal
conductivity, may play an important role in the description of phoretic phenomena.  
A list of measurements concerning thermophoresis on spherical particles can
be found in the review by \cite{You02}, while a critical review on the drag force on a
sphere in the transition regime which includes experimental data 
is given by \cite{Reese}. In the present work,
 a comparison with the more recent data on thermophoresis provided by \cite{Bos02} in
case of a copper sphere in argon gas is presented.  

\section{Formulation of the problem}

\begin{figure}
\centering
\includegraphics[scale=1]{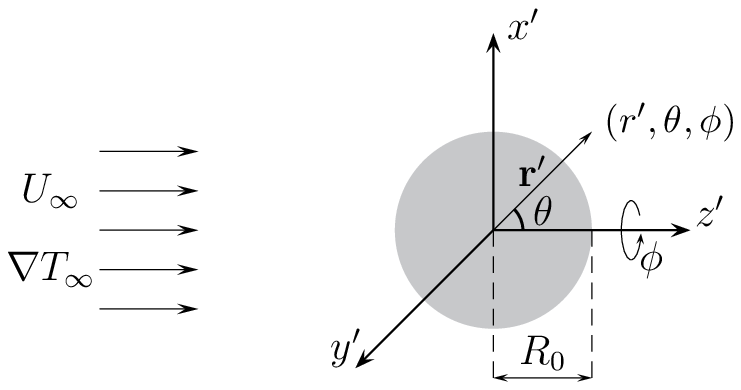}
\caption{Formulation of the problem}
\label{figstat}
\end{figure}

It is considered a sphere of radius $R_0$ at rest placed in a monoatomic
rarefied gas. Far from the sphere, the gas flows with a constant bulk velocity $U_{\infty}$ and
has a temperature gradient $\nabla T_{\infty}$=$\partial T/\partial z'$ in the $z'$-direction as shown in 
figure \ref{figstat}. Due to the problem geometry, it is convenient to introduce 
spherical coordinates $(r',\theta,\phi)$ in the physical space. Then, according to
figure \ref{figstat}, the components of 
the position vector ${\bf r}'$ of gas molecules are given as
\begin{subequations}
\begin{align}
&x'=r'\sin{\theta}\cos{\phi},\\
&y'=r'\sin{\theta}\sin{\phi},\\
&z'=r'\cos{\theta}.
\end{align}
\lae{geo1c}
\end{subequations}
Moreover, the components of the molecular velocity vector ${\bf v}$ read
\begin{subequations}
\begin{align}
&v_x=(v_r\sin{\theta}+v_{\theta}\cos{\theta})\cos{\phi}-v_{\phi}\sin{\phi},\\
&v_y=(v_r\sin{\theta}+v_{\theta}\cos{\theta})\sin{\phi}+v_{\phi}\cos{\phi},\\
&v_z=v_r\cos{\theta}-v_{\theta}\sin{\theta},
\end{align}
\lae{geo2}
\end{subequations} 
where $v_r$, $v_{\theta}$ and $v_{\phi}$ are the radial, polar and azimuthal
components of the molecular velocity vector, respectively, which are written
in spherical coordinates $(v,\theta',\phi')$ in the velocity space as follows
\begin{subequations}
\begin{align}
&v_r=v\cos{\theta'},\\
&v_{\theta}=v_t\cos{\phi'},\\
&v_{\phi}=v_t\sin{\phi'},
\end{align}
\lae{geo3}
\end{subequations}
with the tangential component given as
\begin{equation}
v_t=\sqrt{v_{\theta}^2+v_{\phi}^2}=v\sin{\theta'}.
\lae{geo4}
\end{equation}

It is assumed that the thermal conductivity of the spherical particle is
significantly higher than that corresponding to the carrier gas. As a
consequence, the temperature of the sphere is uniform and equal to the gas temperature in equilibrium
far from the sphere. Let us denote by $n_0$, $T_0$ and $p_0$ the number density,
temperature and pressure of the gas in thermodynamic equilibrium, respectively. 
Two dimensionless thermodynamic forces are introduced here as
follows
\begin{equation}
X_u=\frac{U_{\infty}}{v_0},\quad X_T=\frac{\ell_0}{T_0}\frac{\partial
T}{\partial z'},
\lae{a1}
\end{equation}
where $\ell_0$ and $v_0$ denote the equivalent free path and the most
probable molecular velocity, defined as
\begin{equation}
\quad v_0=\sqrt{\frac{2kT_0}{m}},\quad
\ell_0=\frac{\mu_0v_0}{p_0}.
\lae{a1a}
\end{equation}
 $\mu_0$ denotes the viscosity of the gas at temperature
$T_0$, while $m$ and $k$ are the molecular mass and the Boltzmann constant. 
It is assumed the smallness of the thermodynamic forces defined in
(\ref{a1}), i.e. 
\begin{equation}
|X_u| \ll 1, \quad |X_T| \ll 1.
\lae{a2}
\end{equation}
These assumptions of weak disturbance from equilibrium allow us to split
the problem into two independent parts corresponding to viscous drag and
thermophoresis on the sphere. Hereafter, the dimensionless radius and also the position
and molecular velocity vectors are introduced as follows
\begin{equation}
r_0=\frac{R_0}{\ell_0},\quad {\bf r}=\frac{{\bf r}'}{\ell_0},\quad {\bf
c}=\frac{{\bf v}}{v_0}.
\lae{a3}
\end{equation}
The pressure of the gas is constant and given by the state equation of an
ideal gas as $p_0$=$n_0kT_0$. As a consequence, the asymptotic
behavior of the number density $n({\bf r})$ and temperature $T({\bf r})$ of
the gas far from the sphere are given as
\begin{subequations}
\begin{align}
&n_{\infty}=\lim_{r\rightarrow \infty} n({\bf r})=n_0(1-zX_T),\\
&T_{\infty}=\lim_{r\rightarrow \infty}T({\bf r})=T_0(1+zX_T).
\end{align}
\lae{a6}
\end{subequations}
The main parameter of the problem is the rarefaction parameter, $\delta$,
which is inversely proportional to the Knudsen number, but defined as the ratio 
of the sphere radius to the equivalent molecular free path, i.e.
\begin{equation}
\delta=\frac{R_0}{\ell_0}=r_0.
\lae{a7}
\end{equation}
When $\delta \ll 1$ the gas is in the free molecular regime, while the
opposite limit, $\delta \gg 1$, corresponds to the continuum or hydrodynamic regime. In other
situations, the gas is in the so called transition regime.

The model of gas-surface interaction law proposed by \cite{Cer11} is
employed in the boundary condition. According to this model, the type of the
gas-surface interaction is chosen by setting appropriate values of NEAC
and TMAC. Henceforth, these accommodation coefficients will be denoted by
$\alpha_n$ and $\alpha_t$, respectively. The diffuse scattering or complete
accommodation on the surface corresponds to $\alpha_n$=1 and $\alpha_t$=1.  

The viscous drag and thermophoretic forces acting on the sphere are
calculated in a wide range of the gas rarefaction parameter, $\delta$, so
that all the regimes of the gas flow are covered. Moreover, various values
of accommodation coefficients are considered in the calculations in order to 
analyse the influence of the gas-surface interaction law on the solution of the problem.
The flow fields, i.e. the profiles of the 
density and temperature deviations from equilibrium, bulk velocity and
heat flux, as functions of the radial distance from the sphere are also obtained. 
Some numerical results are compared to those
found in the literature. The reciprocity relation between cross
phenomena is obtained at arbitrary distance from the sphere and then
verified numerically.   

\section{Kinetic equation}

The problem is solved on the basis of the Boltzmann equation for arbitrary
values of rarefaction parameter $\delta$ so that the free molecular, transition
and continuum regimes are covered. For the problem in question, the Boltzmann equation
in the absence of external forces reads
\begin{equation}
{\bf v}\cdot \frac{\partial f}{\partial {\bf r}'}=Q(ff_*),
\lae{ke0}
\end{equation}
where $f$=$f({\bf r}',{\bf v})$ is the distribution function of molecular
velocities and $Q(ff_*)$ is the collision integral whose expression can be found in
the literature, see e.g. \cite{Fer02,CerB2,Kog01}. Here, the model proposed by \cite{Shk02} for the
collision integral is employed due to its reliability to deal with problems
regarding both mass and heat transfer. Then, the collision integral reads
\begin{equation}
Q(ff_*)=Q_S=\nu_S\left\{f^M\left[1+\frac{4}{15}\left(\frac{V^2}{v_0^2}-\frac 52\right)
\frac{{\bf Q}\cdot {\bf V}}{p_0v_0^2}\right]-f({\bf r}',{\bf v}) \right\}
\lae{ke4}
\end{equation}
where $f^M$ is the local Maxwellian function, ${\bf U}({\bf r}')$ and ${\bf
Q}({\bf r}')$ are
the bulk velocity and heat flux vectors. The quantity $\nu_S$ has the order
of the intermolecular interaction frequency and ${\bf V}$=${\bf v}-{\bf U}$
is the so called peculiar velocity. 

The assumptions of smallness of the thermodynamic forces, given in
(\ref{a2}), allow us to linearize the kinetic equation by representing the
distribution function as follows
\begin{equation}
f({\bf r},{\bf c})=f_R^{M}[1+h^{(T)}({\bf r},{\bf v})X_T +
h^{(u)}({\bf r},{\bf v})X_u],
\lae{ke8}
\end{equation}
where $h^{(T)}$ and $h^{(u)}$ are the perturbation functions due to the
thermodynamic forces $X_T$ and $X_u$. The reference Maxwellian function is
given by
\begin{equation}
f_R^{M}=f^{M}=f_0\left[1+z\left(c^2-\frac
52\right)X_T+2c_zX_u\right],
\lae{ke9}
\end{equation}
where $f_0$ is the global Maxwellian function.

Then, after introducing the representation (\ref{ke8}) into (\ref{ke0}), and
also introducing the dimensionless quantities given by (\ref{a3}), the linearized kinetic equation corresponding to each
thermodynamic force is written as
\begin{equation}
\hat{D}h^{(n)}=\hat{L}_Sh^{(n)} + g^{(n)}({\bf r},{\bf c}),\quad n=T, u,
\lae{ke9a}
\end{equation}
where the operator 
\begin{equation}
\hat{D}={\bf c}\cdot\frac{\partial}{\partial {\bf r}}
\end{equation}
and the linearized collision integral reads
\begin{equation}
\hat{L}_Sh^{(n)}
=\nu^{(n)}+\left(c^2-\frac
52\right)\tau^{(n)}+2{\bf c}\cdot{\bf u}^{(n)}+\frac{4}{15}\left(c^2-\frac
52\right){\bf c}\cdot {\bf q}^{(n)}-h^{(n)}.
\lae{ke9a1}
\end{equation}
The free terms are given by
\begin{equation}
g^{(T)}=-c_z\left(c^2-\frac 52\right)\quad g^{(u)}=0.
\lae{ke11}
\end{equation}

The dimensionless quantities in the right hand side of (\ref{ke9a1}) correspond to 
the density and temperature deviations from equilibrium,
bulk velocity and heat flux vectors, respectively, due to the corresponding
thermodynamic force. These quantities are calculated in terms of the
distribution function of molecular velocities, and details regarding these 
calculations are given by \cite{Fer02}. In our notation, these quantities
are written in terms of the perturbation function $h^{(n)}$ corresponding to
each thermodynamic force as 
\begin{equation}
\nu^{(n)}({\bf r})=\frac{1}{\pi^{3/2}}\int h^{(n)}({\bf r},{\bf c})\mbox{e}^{-c^2}\,
\mbox{d}{\bf c},
\lae{ke11a}
\end{equation}
\begin{equation}
\tau^{(n)}({\bf
r})=\frac{2}{3\pi^{3/2}}\int\left(c^2-\frac
32\right)h^{(n)}({\bf r},{\bf c})\mbox{e}^{-c^2}\, \mbox{d}{\bf c},
\lae{ke11b}
\end{equation}
\begin{equation}
{\bf u}^{(n)}({\bf r})=\frac{1}{\pi^{3/2}}\int {\bf
c}h^{(n)}({\bf r},{\bf c})\mbox{e}^{-c^2}\, \mbox{d}{\bf c},
\lae{ke11c}
\end{equation}
\begin{equation}
{\bf q}^{(n)}({\bf r})=\frac{1}{\pi^{3/2}}\int{\bf
c}\left(c^2-\frac 52\right)h^{(n)}({\bf r},{\bf c})\mbox{e}^{-c^2}\,
\mbox{d}{\bf c}.
\lae{ke11d}
\end{equation}

Far from the sphere, i.e. in the limit ${\bf r}\rightarrow \infty$, the asymptotic
behavior of the perturbation functions are obtained from the Chapmann-Enskog
solution for the linearized kinetic equation as
\begin{equation}
h_{\infty}^{(T)}=\lim_{{\bf r}\rightarrow \infty}h^{(T)}({\bf r},{\bf c})=-\frac 32
c_z\left(c^2-\frac 52\right),
\lae{ke11f}
\end{equation}
\begin{equation}
h_{\infty}^{(u)}=\lim_{{\bf r}\rightarrow \infty}h^{(u)}({\bf r},{\bf c})=0.
\lae{ke11e}
\end{equation}

Due to the spherical geometry of the problem, it is convenient to write the kinetic equation
(\ref{ke9a}) in spherical coordinates in both physical and
molecular velocity spaces. Details regarding this transformation are given by
\cite{Shk14}.
Moreover, the problem has simmetry on the azimuthal angle $\phi$. 
Therefore, after some algebraic manipulation, the left hand side of the 
kinetic equation (\ref{ke9a}) is written as follows 
\begin{equation}
\hat{D}h^{(n)}=
c_r\frac{\partial h^{(n)}}{\partial r}- \frac{c_t}{r}\frac{\partial
h^{(n)}}{\partial \theta'}+\frac{c_t}{r}\cos{\phi'}\frac{\partial
h^{(n)}}{\partial \theta}-
\frac{c_t}{r}\sin{\phi'}\cot{\theta}\frac{\partial
h^{(n)}}{\partial \phi'}
\lae{ke1}
\end{equation}
where $h^{(n)}$=$h^{(n)}(r,\theta,{\bf c})$ and ${\bf c}$=$(c, \theta',\phi')$.
The simmetry of the solution on the azimuthal angle also allows us to eliminate the dependence of the moments 
of the perturbation function which appear in the right hand side of the
kinetic equation on the angle $\phi$. Thus, the density and temperature deviations given in
(\ref{ke11a}) and (\ref{ke11b}) are written as 
\begin{equation}
\nu^{(n)}(r,\theta)=\frac{1}{\pi^{3/2}}\int
h^{(n)}(r,\theta,{\bf c})\mbox{e}^{-c^2}\, \mbox{d}{\bf c}
\lae{ke3a}
\end{equation}
\begin{equation}
\tau^{(n)}(r,\theta)=\frac{2}{3\pi^{3/2}}\int \left(c^2-\frac
32\right)h^{(n)}(r,\theta,{\bf c})\mbox{e}^{-c^2}\,
\mbox{d}{\bf c},
\lae{ke3b}
\end{equation}
where $\mbox{d}{\bf c}$=$c^2\sin{\theta'}\mbox{d}c\mbox{d}\theta'\mbox{d}\phi'$.
Moreover, the nonzero components of the bulk velocity and heat flux vectors
given in (\ref{ke11c}) and (\ref{ke11d}) are written as
\begin{equation}
u_r^{(n)}(r,\theta)=\frac{1}{\pi^{3/2}}\int
c_rh^{(n)}(r,\theta,{\bf c})\mbox{e}^{-c^2}\, \mbox{d}{\bf
c},
\lae{ke3c}
\end{equation}
\begin{equation}
u_{\theta}^{(n)}(r,\theta)=\frac{1}{\pi^{3/2}}\int
c_{\theta}h^{(n)}(r,\theta,{\bf c})\mbox{e}^{-c^2}\, \mbox{d}{\bf
c},
\lae{ke3d}
\end{equation}
\begin{equation}
q_r^{(n)}(r,\theta)=\frac{1}{\pi^{3/2}}\int c_r\left(c^2-\frac
52\right)h^{(n)}(r,\theta,{\bf c})\mbox{e}^{-c^2}\,
\mbox{d}{\bf c},
\lae{ke3e}
\end{equation}
\begin{equation}
q_{\theta}^{(n)}(r,\theta)=\frac{1}{\pi^{3/2}}\int c_{\theta}\left(c^2-\frac
52\right)h^{(n)}(r,\theta,{\bf c})\mbox{e}^{-c^2}\,
\mbox{d}{\bf c}.
\lae{ke3f}
\end{equation}

Similarly to the moments appearing in the kinetic equation, the force on the sphere in the $z'$-direction is
 calculated in terms of the distribution function of molecular velocities as
follows
\begin{equation}
F_z'=-\int_{\Sigma'_w}\mbox{d}\Sigma'_w\int mv_rv_z f({\bf r},{\bf v})\,
\mbox{d}{\bf v}
\lae{ke3g}
\end{equation}
where $\mbox{d}\Sigma'_w$=$R_0^2\sin{\theta}\mbox{d}\theta\mbox{d}\phi$ is
an area element taken in the surface of the sphere. For convenience, the
dimensionless force is introduced here as
\begin{equation}
F_z=\frac{F_z'}{4\pi R_0^2 p_0}.
\lae{ke3ga}
\end{equation}
Then, after the introduction of the representation (\ref{ke8}) into (\ref{ke3g})
and some algebraic manipulation, the dimensionless force acting on the sphere reads
\begin{equation}
F_z=F_TX_T + F_uX_u,
\lae{ke3gb}
\end{equation}
where the dimensionless thermophoretic and drag forces are given as
\begin{equation}
F_{\mbox{\tiny{T}}}=
-\frac{1}{2\pi^{5/2}}\int_{\Sigma_w}\mbox{d}\Sigma_w\int c_rc_z\mbox{e}^{-c^2}\left[h^{(T)}+z_0\left(c^2-\frac
52\right)\right]\, \mbox{d}{\bf c},
\lae{ke3j}
\end{equation}
\begin{equation}
F_u=-\frac{1}{2\pi^{5/2}}\int_{\Sigma_w}\mbox{d}\Sigma_w
\int c_rc_z\mbox{e}^{-c^2}(h^{(u)}+2c_z)\, \mbox{d}{\bf c},
\lae{ke3i}
\end{equation}
with $\mbox{d}\Sigma_w$=$\mbox{d}\Sigma'_w/R_0^2$.

\section{Boundary condition}

The boundary conditions for both the drag and the thermophoresis on
the sphere are obtained from the relation between the disbribution functions
of incident particles on the wall and reflected particles from the wall. 
According to \cite{CerB2,Sha02B}, the general form of the linearized boundary
condition at the surface reads
\begin{equation}
h^{+(n)}=\hat{A}h^{-(n)}+h_w^{(n)}-\hat{A}h_w^{(n)},
\lae{bc1}
\end{equation}
where the signal $+$ denotes the reflected particles from the surface, while
the signal $-$ denotes the incident particles on the surface. From
(\ref{ke9}), the source terms are given as
\begin{equation}
h_w^{(u)}=-2c_z,\quad h_w^{(T)}=-z_0\left(c^2-\frac 52\right),
\lae{bc2}
\end{equation}
where $z_0$=$r_0\cos{\theta}$ and
$c_z$=$c_r\cos{\theta}-c_{\theta}\sin{\theta}$.

For the problem in question, the scattering operator $\hat{A}$ is expressed as
\begin{equation}
\hat{A}h^{(n)}=\hat{A}_r\hat{A}_{\theta}\hat{A}_{\phi}h^{(n)},
\lae{bc2a}
\end{equation}
where
\begin{equation}
\hat{A}_r\xi=
-\frac{1}{c_r}\int_{c_r'<0}c_r'\exp{(c_r^2-c_r'^{2})}R_r(
c_r\rightarrow c_r')\xi\,\mbox{d}c_r',
\lae{bc3}
\end{equation}
\begin{equation}
\hat{A}_i\xi=\int_{-\infty}^{\infty}\exp{(c_i^2-c_i'^2)}R_i(c_i\rightarrow
c_i')\xi \, \mbox{d}c_i',\quad i=\theta, \phi,
\lae{bc3a}
\end{equation}
for an arbitrary $\xi$ as function of the molecular velocity. The scattering kernel
proposed by Cercignani and Lampis is decomposed as follows
\begin{equation}
R({\bf c}\rightarrow {\bf c}')=R_r(c_r\rightarrow c_r')R_{\theta}(c_{\theta}
\rightarrow c_{\theta}')R_{\phi}(c_{\phi}\rightarrow c_{\phi}'),
\lae{bc4}
\end{equation}
where
\begin{equation}
R_r(c_r\rightarrow
c_r')=\frac{2c_r}{\alpha_n}\exp{\left[-\frac{c_r^2+(1-\alpha_n)c_r'^2}{\alpha_n}\right]}
I_0\left(\frac{2\sqrt{1-\alpha_n}}{\alpha_n}c_rc_r'\right),
\lae{bc5}
\end{equation}
\begin{equation}
R_i(c_i\rightarrow c_i')=\frac{1}{\sqrt{\pi
\alpha_t(2-\alpha_t)}}\exp{\left\{-\frac{[c_i-(1-\alpha_t)
c_i']^2}{\alpha_t(2-\alpha_t)}\right\}},\quad i=\theta, \phi.
\lae{bc6}
\end{equation}
$I_0$ denotes the modified Bessel function of first kind and zeroth
order given by \cite{Abr01}. According to this model, the accommodation coefficients can vary in the
ranges $0 \le \alpha_t \le 2$ and $0\le \alpha_n \le 1$. The case 
$\alpha_t$=1 and $\alpha_n$=1 corresponds to diffuse scattering or complete
accommodation on the spherical surface, while the case $\alpha_t$=0 and $\alpha_n$=0
corresponds to specular reflection at the surface. It is worth noting that,
for intermediate values of $\alpha_t$ and $\alpha_n$ the scattering kernel
proposed by Cercignani-Lampis differs significantly from the diffuse-specular model proposed
by Maxwell, in which just one accommodation coefficient was introduced.    

It can be shown that the following relations are satisfied
\begin{equation}
\hat{A}_ic_i=(1-\alpha_t)c_i,\quad i=\theta, \phi,
\lae{bc7}
\end{equation}
\begin{equation}
\hat{A}_ic_i^2=(1-\alpha_t)^2c_i^2+\frac 12 \alpha_t(2-\alpha_t),
\lae{bc8}
\end{equation}
\begin{equation}
\hat{A}_ic_i^3=(1-\alpha_t)^3c_i^3+\frac 32
\alpha_t(2-\alpha_t)(1-\alpha_t)c_i,
\lae{bc9}
\end{equation}
\begin{equation}
\hat{A}_rc_r=-\sqrt{\alpha_n}H_1(\eta),
\lae{bc10}
\end{equation}
\begin{equation}
\hat{A}_rc_r^2=\alpha_n + (1-\alpha_n)c_r^2,
\lae{bc11}
\end{equation}
\begin{equation}
\hat{A}_rc_r^3=-\alpha_n^{3/2}H_3(\eta),
\lae{bc12}
\end{equation}
where
\begin{equation}
H_j(\eta)=2\mbox{e}^{-\eta^2}\int_{0}^{\infty}\xi^{j+1}\mbox{e}^{-\xi^2}I_0(2\eta
\xi)\, \mbox{d}\xi,\quad \xi=\frac{c_r'}{\sqrt{\alpha_n}},\quad j=1, 3,
\lae{bc13}
\end{equation}
and 
\begin{equation}
\eta=c_r\sqrt{\frac{1}{\alpha_n}-1}.
\lae{bc12a}
\end{equation}

Therefore, with the help of the relations (\ref{bc7})-(\ref{bc12}), the boundary conditions
at $r$=$r_0$ and $c_r > 0$, for each
thermodynamic force are obtained from (\ref{bc1}) as 
\begin{equation}
h^{+(T)}=\hat{A}h^{-(T)}+z_0[\alpha_n(1-c_r^2)+\alpha_t(2-\alpha_t)(1-c_t^2)],
\lae{bc15}
\end{equation}
\begin{equation}
h^{+(u)}=\hat{A}h^{-(u)}-2\frac{z_0}{\delta}[(1-\alpha_t)
c_r+\sqrt{\alpha_n}H_1(\eta)]-2\alpha_tc_z.
\lae{bc14}
\end{equation}

\section{Reciprocity relation}

As it is known from the non-equilibrium thermodynamics, see e.g.
\cite{Deg01}, the reciprocity relations between cross phenomena represent
an important criterion to verify the numerical precision in calculations
regarding small deviations from thermodynamic equilibrium. According to
\cite{Sha60,Sha62,Sha81}, the reciprocity relation for the problem in
question can be written as
\begin{equation}
\Lambda_{uT}^t=\Lambda_{Tu}^t,
\lae{rec1}
\end{equation}
where the time-reversal kinetic coefficients are defined as
\begin{equation}
\Lambda_{kn}^t=((\hat{T}g'^{(k)},h^{(n)}))+\int_{\Sigma_w}(\hat{T}v_rh_w^{(k)},h^{(n)})\,
\mbox{d}\Sigma
+\frac 12 \int_{\Sigma_g}(\hat{T}v_rh^{(k)},h^{(n)})\, \mbox{d}\Sigma.
\lae{rec2}
\end{equation}
The dimension free terms $g'^{(n)}$=$v_0g^{(n)}/\ell_0$ ($n=u, T$), where $g^{(n)}$ 
are given in (\ref{ke11}). The source terms $h_w^{(n)}$ are given in
(\ref{bc2}). Here, the time reversal
operator $\hat{T}$ just changes the sign of the molecular velocity, i.e.
$\hat{T}h({\bf v})$=$h(-{\bf v})$. The scalar products are defined as
\begin{equation}
(\xi_1,\xi_2)=\int f_0\xi_1({\bf r}',{\bf v})\xi_2({\bf r}',{\bf v})\,
\mbox{d}{\bf v},
\lae{rec3}
\end{equation}
and
\begin{equation}
((g,h))=\int_{\Omega}(g,h)\, \mbox{d}{\bf r}'.
\lae{rec4}
\end{equation}
$\Omega$ means the gas flow domain, while $\Sigma_w$ and $\Sigma_g$ mean the
solid spherical surface and the imaginary spherical surface at $r'
\rightarrow \infty$ which enclose the gas domain.

Then, after some algebraic manipulation, the time reversed kinetic
coefficients are written as
\begin{equation}
\Lambda_{uT}^t=-4\pi R_0^2 n_0v_0F_T - \frac 12
\int_{\Sigma_g}(\hat{T}v_rh^{(T)},h^{(u)})\, \mbox{d}\Sigma,
\lae{rec5}
\end{equation}
\begin{equation}
\Lambda_{Tu}^t=v_0n_0\int_{\Sigma_g}zq_r^{(u)}(r,\theta)\, \mbox{d}\Sigma +
\frac 12 \int_{\Sigma_g}(\hat{T}v_rh^{(T)},h^{(u)})\, \mbox{d}\Sigma.
\lae{rec6}
\end{equation}
Therefore, after substituting (\ref{rec5}) and (\ref{rec6}) into
(\ref{rec1}), the thermophoretic force on the sphere is related to the
solution of the drag force problem as
\[
F_{\mbox{\tiny{T}}}=-\frac{r^2}{2\delta^2}\biggl[r\int_{0}^{\pi}q_r^{(u)}(r,\theta)\cos{\theta}\sin{\theta}\,
\mbox{d}\theta
\]
\begin{equation}
\hskip2cm
+\frac{1}{\pi^{3/2}}\int_{0}^{\pi}\int c_rh^{(T)}(r,\theta,-{\bf
c})h^{(u)}(r, \theta,{\bf c})\mbox{e}^{-c^2}\sin{\theta}\,\mbox{d}{\bf
c}\mbox{d}\theta\biggl],
\lae{rec7}
\end{equation}
where $r$ is the radius of the imaginary spherical surface $\Sigma_g$, which
can be arbitrary. The right hand side of the relation (\ref{rec7}) was
calculated numerically for $r$=0, 5, 10 and 40 and the fulfillment of such
a relation was verified numerically within the numerical error of 0.1\%.

\section{Method of solution}

\subsection{Free molecular regime}

In the free molecular regime, i.e. $\delta << 1$, the
collision integral which appears in the Boltzmann equation (\ref{ke0}) can be
neglected. As a consequence, in this regime of the gas flow, the problem is solved analytically via 
solution of a differential equation for each thermodynamic force obtained from (\ref{ke9a}) as
\begin{equation}
\hat{D} h^{(n)}=g^{(n)},
\lae{fm1}
\end{equation} 
whose solutions must satisfy the boundary
conditions given in (\ref{bc1}). Moreover, in
this regime of the gas flow,  the distribution function of incident gas particles on the
surface is not perturbed, which means that $h^{-(n)}$=$h_{\infty}^{(n)}$ 
as given by (\ref{ke11f}) and (\ref{ke11e}).   
The method of the characteristics allows
us to solve the previous equations for each thermodynamic force and, thus, obtain the following
solutions
\begin{equation}
h^{(n)}(r,\theta,{\bf c})=
\begin{cases}
\begin{split}
&h_c^{(n)}\cos{\theta}+h_s^{(n)}c_{\theta}\sin{\theta}+g^{(n)}\frac Sc,\quad 0\le \theta' \le
\theta_0,\\[0.25cm]
& h_{\infty}^{(n)},\quad \theta_0 \le \theta'\le \pi,
\end{split}
\end{cases}
\lae{fm2}
\end{equation}
where
\begin{equation}
h_c^{(n)}(r,c,\theta')=C_1^{(n)}\left(r-\frac Sc c_r\right) - C_2^{(n)}c_r,
\lae{fm2a}
\end{equation}
\begin{equation}
h_s^{(n)}(r,c,\theta')=C_1^{(n)}\frac Sc + C_2^{(n)}.
\lae{fm2b}
\end{equation}
$S$ is the distance between a point in the gas flow domain with 
Cartesian coordinates $(x,y,z)$ and a point in the spherical surface 
with Cartesian coordinates $(x_0,y_0,z_0)$ which is written as 
\begin{equation}
S=r\cos{\theta'}-\sqrt{r_0^2-r^2\sin^2{\theta'}}.
\lae{fm4}
\end{equation}
It is worth noting that the vector ${\bf S}$ is directed towards -${\bf c}$
and, consequently, the following relation is valid 
\begin{equation}
z_0=z-\frac Sc c_z.
\lae{fm4a}
\end{equation}
The angle $\theta_0$ is given by
\begin{equation}
\theta_0=\arcsin{\left(\frac{r_0}{r}\right)}.
\lae{fm5}
\end{equation}
The quantities $C_1^{(n)}$ and $C_2^{(n)}$ are obtained from the boundary
conditions given in (\ref{bc15}) and (\ref{bc14}) as
\[
C_1^{(T)}=-\frac
{3}{2r_0}\left\{\alpha_n^{3/2}H_3(\eta)+\alpha_n^{1/2}H_1(\eta)\left[
\alpha_t(2-\alpha_t)+(1-\alpha_t)^2c_t^2-\frac 52 \right] \right\}
\]
\begin{equation}
\hskip2cm
+\alpha_n(1-c_r^2)+\alpha_t(2-\alpha_t)-\alpha_t(2-\alpha_t)c_t^2
+\frac{2\alpha_t c_r}{r_0}
\lae{fm5c}
\end{equation}
\begin{equation}
C_2^{(T)}=\frac 32 (1-\alpha_t)\left[\alpha_n
+(1-\alpha_n)c_r^2+(1-\alpha_t)^2c_t^2+2\alpha_t(2-\alpha_t)-\frac
52\right].
\lae{fm5d}
\end{equation}
\begin{equation}
C_1^{(u)}=-\frac{2[(1-\alpha_t)c_r+\sqrt{\alpha_n}H_1(\eta)]}{r_0},
\lae{fm5a}
\end{equation}
\begin{equation}
C_2^{(u)}=2\alpha_t,
\lae{fm5b}
\end{equation}
where $H_1(\eta)$ and $H_3(\eta)$ are defined in (\ref{bc13}).

Then, the thermophoretic and drag forces on the sphere are obtained
from (\ref{ke3j}) and (\ref{ke3i}) as
\begin{equation}
F_{\mbox{\tiny{T}}}=-\frac{1}{2\sqrt{\pi}}\left(\frac{1+\alpha_t}{2}+\EuScript{H}\right),
\lae{fm10}
\end{equation}
\begin{equation}
F_u=\frac{2}{3\sqrt{\pi}}\left(1+\alpha_t+2\alpha_n^{1/2}\int_0^{\infty}
c_r^2\mbox{e}^{-c_r^2}
H_1(\eta)\, \mbox{d}c_r\right),
\lae{fm9}
\end{equation}
where
\begin{equation}
\EuScript{H}=
\alpha_n^{1/2}\int_{0}^{\infty}c_r^2\mbox{e}^{-c_r^2}\left[\alpha_nH_3(\eta)-\frac 32
H_1(\eta)\right]\, \mbox{d}c_r,
\lae{fm11}
\end{equation}
with $\eta$ and $\xi$ defined in (\ref{bc13}) and (\ref{bc12a}),
respectively.

In case of diffuse scattering, i.e. $\alpha_t$=1 and $\alpha_n$=1, 
the thermophoretic and drag forces given in
(\ref{fm10}) and (\ref{fm9}) correspond to those found in the literature,
see e.g. \cite{Tak09,Ber09,Son48}.

The macrocroscopic characteristics of the gas flow around the sphere due to each thermodynamic force
 can be obtained just by substituting the corresponding solution given in
(\ref{fm2}) into the expressions (\ref{ke3a})-(\ref{ke3f}).

\subsection{Arbitrary gas rarefaction}

In order to consider arbitrary values of the gas rarefaction, the problem
is solved numerically by employing the linearized kinetic equations given in (\ref{ke1})
for each thermodynamic force subject to the corresponding boundary
condition. Here, these equations are solved by the discrete velocity
method, whose details can be found in the literature, see e.g.
\cite{Sha11,Sha02B}. Moreover, the split method 
proposed by \cite{Nar02} to deal with the problem of the discontinuiy of the
distribution function of molecular velocities is employed. In rarefied gas
dynamics, the problem of the discontinuity of the distribution
function is a peculiarity inherent to gas flows around convex bodies, see
e.g. \cite{Son13}, and must be treated carefully when a finite difference scheme is used.  
 The idea of the split method is the decomposition of the perturbation function into
two parts as
\begin{equation}
h^{(n)}({\bf r},{\bf c})=h_0^{(n)}({\bf r},{\bf c})+ \tilde{h}^{(n)}({\bf
r},{\bf c}),
\lae{split1}
\end{equation}
where the function $h_0^{(n)}$ is obtained from the solution of
the differential equation
\begin{equation}
\hat{D} h_0^{(n)}-h_0^{(n)}=0,
\lae{split2}
\end{equation}  
with boundary conditions
\begin{equation}
h_0^{+(n)}=\hat{A}h_{\infty}^{(n)}+h_w^{(n)}-\hat{A}h_w^{(n)},
\lae{split3}
\end{equation}
where $h_w^{(n)}$ and $h_{\infty}^{(n)}$ are given in (\ref{bc2}),
(\ref{ke11f}) and (\ref{ke11e}). Note that, the discontinuous perturbation
function $h_0^{(n)}$ can be obtained analytically by the characteristics
method. Thus, the solution of equation (\ref{split2}) subject to the boundary condition
(\ref{split3}), is written as follows
\begin{equation}
h_0^{(n)}(r,\theta,{\bf c})=
\begin{cases}
\begin{split}
&[h_c^{(n)}\cos{\theta}+h_s^{(n)}c_{\theta}\sin{\theta}]\mbox{e}^{-S/c},
\quad 0 \le \theta' \le \theta_0,\\
&h_{\infty}^{(n)},\quad \theta_0 < \theta' \le \pi,
\end{split}
\end{cases}
\lae{split4}
\end{equation}
where the distance $S$ along the characteristic line and the angle
$\theta_0$ are given in (\ref{fm4}) and (\ref{fm5}), while the functions
$h_c^{(n)}$ and $h_s^{(n)}$ correspond to those given in (\ref{fm2a}) and (\ref{fm2b})
in the free molecular regime.
Note that the moments (\ref{ke3a})-(\ref{ke3f}) are also decomposed into two parts
as 
\begin{equation}
\Lambda^{(n)}=\Lambda_0^{(n)}+\tilde{\Lambda}^{(n)},\quad
\Lambda^{(n)}=\nu^{(n)}, \tau^{(n)}, u_r^{(n)}, q_r^{(n)}, u_{\theta}^{(n)},
q_{\theta}^{(n)},
\lae{split4a}
\end{equation}
where the quantities with tilde are calculated by the same expressions given
in (\ref{ke3a})-(\ref{ke3f}), just replacing $h^{(n)}$ by $\tilde{h}^{(n)}$,
while the quantities with zero index are calculated in terms of the known
solutions $h_0^{(n)}$ given in (\ref{split4}).

The function $\tilde{h}^{(n)}$ satisfy the kinetic equation (\ref{ke9a})
just replacing $h^{(n)}$ by $\tilde{h}^{(n)}$, but
with the boundary condition 
\begin{equation}
\tilde{h}^{+(n)}=\hat{A}\tilde{h}^{-(n)}.
\lae{split5}
\end{equation}
The advantage of the split method is that the function $\tilde{h}^{(n)}$ is
sufficiently smooth so that a finite difference scheme leads to a smaller
numerical error. To reduce the number of variables of the perturbation
function $\tilde{h}^{(n)}$, its dependence on the variables 
$\theta$ and $\phi'$ is eliminated by employing the similarity solution proposed by
\cite{Son32}. Then, in our notation, the perturbation function
$\tilde{h}^{(n)}$ is represented as
\begin{equation}
\tilde{h}^{(n)}(r,\theta,{\bf c})=\Phi^{(n)}(r,c,\theta')\cos{\theta}+
\Psi^{(n)}(r,c,\theta')c_{\theta}\sin{\theta}.
\lae{ms1}
\end{equation}  
Then, the substitution of the representation (\ref{ms1}) into the kinetic
equation (\ref{ke9a}) allows us to obtain a system
of equations for the new perturbation functions $\Phi^{(n)}$ and $\Psi^{(n)}$
corresponding to each thermodynamic force given as 
\[
c_r\frac{\partial \Phi^{(n)}}{\partial r}-\frac{c_t}{r}\frac{\partial
\Phi^{(n)}}{\partial
\theta'}+\frac{c_t^2}{r}\Psi^{(n)}=\nu^{*(n)}+\left(c^2-\frac
32\right)\tau^{*(n)}+2c_ru_r^{*(n)}
\]
\begin{equation}
\hskip2cm
+\frac{4}{15}c_r\left(c^2-\frac 52 \right)q_r^{*(n)}-\Phi^{(n)}+g_1^{*(n)},
\lae{ms2}
\end{equation}

\[
c_r\frac{\partial \Psi^{(n)}}{\partial r}-\frac{c_t}{r}\frac{\partial
\Psi^{(n)}}{\partial \theta'}-\frac{c_r}{r}\Psi^{(n)}-\frac
1r\Phi^{(n)}=2u_{\theta}^{*(n)}
\]
\begin{equation}
\hskip2cm
+\frac{4}{15}\left(c^2-\frac 52
\right)q_{\theta}^{*(n)}-\Psi^{(n)}+g_2^{*(n)},
\lae{ms3}
\end{equation}
where the free terms read
\begin{equation}
g_1^{*(T)}=-c_r\left(c^2-\frac 52\right),\quad
g_2^{*(T)}=c^2-\frac 52,\quad
g_1^{*(u)}=g_2^{*(u)}=0.
\lae{ms4}
\end{equation}

The dimensionless moments which appear in the right hand side of
equations (\ref{ms2}) and (\ref{ms3}) are obtained from
(\ref{ke3a})-(\ref{ke3f}) as
\begin{equation}
\nu^{*(n)}(r)=\frac{\nu_0^{(n)}(r)}{\cos{\theta}}+
\frac{2}{\sqrt{\pi}}\int_{0}^{\infty}\int_{0}^{\pi}c_t\Phi^{(n)}\mbox{e}^{-c^2}\,
c\mbox{d}c\mbox{d}\theta',
\lae{ms5}
\end{equation}
\begin{equation}
\tau^{*(n)}(r)=\frac{\tau_0^{(n)}(r)}{\cos{\theta}}+
\frac{4}{3\sqrt{\pi}}\int_{0}^{\infty}\int_{0}^{\pi}\left(c^2-\frac
32\right)c_t\Phi^{(n)}\mbox{e}^{-c^2}\,
c\mbox{d}c \mbox{d}\theta',
\lae{ms6}
\end{equation}
\begin{equation}
u_r^{*(n)}(r)=\frac{u_{r0}^{(n)}(r)}{\cos{\theta}} +
\frac{2}{\sqrt{\pi}}\int_{0}^{\infty}\int_{0}^{\pi}c_rc_t
\Phi^{(n)}\mbox{e}^{-c^2}\,
c\mbox{d}c \mbox{d}\theta',
\lae{ms7}
\end{equation}
\begin{equation}
u_{\theta}^{*(n)}(r)=\frac{u_{\theta 0}^{(n)}(r)}{\sin{\theta}} +
\frac{1}{\sqrt{\pi}}\int_{0}^{\infty}\int_{0}^{\pi}
c_t^3\Psi^{(n)}\mbox{e}^{-c^2}\,
c\mbox{d}c \mbox{d}\theta',
\lae{ms8}
\end{equation}
\begin{equation}
q_r^{*(n)}(r)=\frac{q_{r0}^{(n)}(r)}{\cos{\theta}} +
\frac{2}{\sqrt{\pi}}\int_{0}^{\infty}\int_{0}^{\pi}c_rc_t\left(c^2-\frac
52\right)\Phi^{(n)}\mbox{e}^{-c^2}\, c\mbox{d}c \mbox{d}\theta',
\lae{ms9}
\end{equation}
\begin{equation}
q_{\theta}^{*(n)}(r)=\frac{q_{\theta 0}^{(n)}(r)}{\sin{\theta}} +
\frac{1}{\sqrt{\pi}}\int_{0}^{\infty}\int_{0}^{\pi}c_t^3\left(c^2-\frac
52\right)\Psi^{(n)}\mbox{e}^{-c^2}\,
c\mbox{d}c\mbox{d}\theta'.
\lae{ms10}
\end{equation}

Far from the sphere the asymptotic behavior of the functions $\Phi^{(n)}$ and $\Psi^{(n)}$ are
obtained from (\ref{ke11f}) and (\ref{ke11e}) as
\begin{equation}
\Phi_{\infty}^{(T)}=\lim_{r\rightarrow \infty}\Phi^{(T)}=-\frac 32 c_r\left(c^2-\frac
52\right),\quad \Psi_{\infty}^{(T)}=\lim_{r\rightarrow \infty}\Psi^{(T)}=\frac 32\left(c^2-\frac
52\right),
\lae{ms12}
\end{equation}
\begin{equation}
\Phi_{\infty}^{(u)}=\lim_{r\rightarrow \infty} \Phi^{(u)}=0,\quad
\Psi_{\infty}^{(u)}=\lim_{r\rightarrow
\infty}\Psi^{(u)}=0.
\lae{ms11}
\end{equation}

The representation (\ref{ms1}) allows us to write the dimensionless forces
given in (\ref{ke3j}) and (\ref{ke3i}) as
\[
F_n=-\frac{4}{3\sqrt{\pi}}\int_{-\infty}^{\infty}\int_{-\infty}^{\infty}c_r^2c_t\Phi^{(n)}(r_0,c_r,c_t)\mbox{e}^{-c^2}\,
\mbox{d}c_r\mbox{d}c_t
\]
\begin{equation}
\hskip2cm
+\frac{4}{3\sqrt{\pi}}\int_{-\infty}^{\infty}\int_{-\infty}^{\infty}c_rc_t^3\Psi^{(n)}(r_0,c_r,c_t)\mbox{e}^{-c^2}\,
\mbox{d}c_r\mbox{d}c_t.
\lae{ms13}
\end{equation}

Regarding the boundary conditions, the representation (\ref{ms1}) is compatible with the 
Cercignani-Lampis boundary condition taking the following form
\begin{equation}
\hat{A}h^{-(n)}=\cos{\theta}\hat{A}_r\hat{A}_t^{(0)}\Phi^{-(n)}+
c_t\cos{\phi'}\hat{A}_r\hat{A}_t^{(1)}\Psi^{-(n)},
\lae{ms14}
\end{equation}
where
\[
\hat{A}_r\xi=\frac{2}{\alpha_n}\int_{0}^{\infty}c_r'\exp{\left[-\frac{(1-\alpha_n)c_r^2+c_r'^2}{\alpha_n}\right]}
\]
\begin{equation}
\hskip2cm
\times
I_0\left(\frac{2\sqrt{1-\alpha_n}c_rc_r'}{\alpha_n}\right)\xi(-c_r',c_t')\,\mbox{d}c_r',
\lae{ms15}
\end{equation}
\[
\hat{A}_t^{(i)}\xi=\frac{2}{\alpha_t(2-\alpha_t)}\int_{0}^{\infty}c_t'\exp{\left[-\frac{(1-\alpha_t)^2c_t^2+c_t'^2}
{\alpha_t(2-\alpha_t)}\right]}
\]
\begin{equation}
\hskip2cm
\times I_i\left[\frac{2(1-\alpha_t)c_tc_t'}{\alpha_t(2-\alpha_t)}\right]
\xi(c_r',c_t')\, \mbox{d}c_t',
\lae{ms16}
\end{equation}
where $I_i$ ($i$=0, 1) is the modified Bessel function of first kind and
$i$-th order.

Therefore, the boundary conditions for the perturbation functions $\Phi^{(n)}$ and $\Psi^{(n)}$ are
obtained from (\ref{bc15}) and (\ref{bc14}) as follows
\begin{equation}
\Phi^{+(n)}=\hat{A}_r\hat{A}_t^{(0)}\Phi^{-(n)},\quad 
\Psi^{+(n)}=\hat{A}_r\hat{A}_t^{(1)}\Psi^{-(n)}.
\lae{ms19}
\end{equation}

Then, the system of kinetic equations (\ref{ms2}) and (\ref{ms3}) for each
thermodynamic force, subject to the corresponding conditions
(\ref{ms12})-(\ref{ms11}) and
(\ref{ms19}), were solved numerically via the discrete velocity method with an accuracy of 0.1\% for the
moments of the perturbation functions. Such an accuracy was estimated by
varying the grid parameters $N_r$, $N_{c}$ and $N_{\theta}$ corresponding to
the number of nodes in the radial coordinate $r$, molecular
speed $c$ and angle $\theta'$. Moreover, the reciprocal relation \ref{rec7}
was verified.

\section{Results and discussion}

\subsection{Free molecular regime}

Firstly, the results obtained from the analytic solutions (\ref{fm10}) and
(\ref{fm9}) were compared to those given by \cite{Che25} in the free molecular
regime. Figures (\ref{figCher1}) and (\ref{figCher2}) show the profiles of the
dimensionless thermophoretic and drag forces 
on the sphere as functions of the TMAC, $\alpha_t$, and fixed values of
NEAC corresponding to $\alpha_n$= 0.1, 0.5 and 0.9. Figures
(\ref{figCher3}) and (\ref{figCher4}) show the forces on the sphere as functions of the
NEAC and fixed values of TMAC corresponding to $\alpha_t$=0, 0.4, 0.8 and 1. 
The analytic expressions given by \cite{Che25}
to calculate these forces were obtained in the limit $(1-\alpha_t)\ll 1$ and
$(1-\alpha_n)\ll 1$, which means almost complete accommodation of gas
particles on the sphere.  According to figures (\ref{figCher1}) and
(\ref{figCher3}), there is a good
agreement between the results obtained in the present work and those given by
\cite{Che25} for the drag force in the whole range of accommodation
coefficients. There is a small difference between the results only for
small values of NEAC. On the contrary, as
one can see from figures (\ref{figCher2}) and (\ref{figCher4}), there is a
large disagreement between the present results and
those given by \cite{Che25} for the thermophoretic force in the whole
range of accommodation coefficients. Note that, even in the limit of diffuse scattering 
such a disagreement is still large. However, it was verified numerically that
the reciprocity relation given by (\ref{rec7}) is fullfilled within an accuracy of 0.1\% 
for arbitrary values of accommodation coefficients. Table \ref{tabCA} presents the
comparison between the values of the thermophoretic force on the sphere
calculated by (\ref{rec7}) and (\ref{fm10}). Moreover, 
numerical results for both forces in the free molecular
regime are given in tables \ref{tabA} and \ref{tabB}. As one can see from
these tables, the numerical results obtained via kinetic equation for small value of
rarefaction parameter tend to those given by (\ref{fm10}) and (\ref{fm9}).
Therefore, since the reciprocity relation between cross 
phenomena was not verified by \cite{Che25}, the disagreement between the results for the 
thermophoretic force should be explained by some error in the derivation of the analytic expression
presented by these authors. In case of the diffuse scattering, (\ref{fm10}) and (\ref{fm9}) lead to 
the following expressions for the thermophoretic and drag forces
\begin{equation}
F_{\mbox{\tiny{T}}}=-\frac{1}{2\sqrt{\pi}},
\lae{res1}
\end{equation}
\begin{equation}
F_u=\frac{4}{3\sqrt{\pi}}\left(1+\frac{\pi}{8}\right),
\lae{res2}
\end{equation}
which are well known from the literature, e.g. \cite{Tak09,Ber09,Son48}.

\subsection{Transitional regime}

Under the assumption of complete accommodation of gas particles on the
surface, the results obtained for the thermophoretic and viscous drag forces were
compared to those presented by \cite{Ber09,Ber10,Tak07,Tak09}. The comparison is shown
in figures \ref{figA} and \ref{figB}, in which $F_{\mbox{\tiny{T}}}^*$ and $F_u^*$ denote
the ratio of the thermophoretic and drag forces to the
corresponding values (\ref{res1}) and (\ref{res2}) in the free molecular
regime. \cite{Ber09,Ber10}k
used the integral-moment method to solve the same linearized kinetic equation
of the present work, while \cite{Tak07, Tak09} solved the full linearized Boltzmann
equation via a finite difference scheme method and the similarity
solution proposed by \cite{Son02}. It is worth noting
that the integral-moment method consists on obtaining a set of integral
equations for the moments of the distribution function and its advantage is
that only the physical space must be discretized. 
However, this method requires much more
computational memory and CPU time than that required when the discrete
velocity method is employed. Regarding the solution of the full 
Boltzmann equation, in spite of the great computational infrastructure
currently available, to find this solution is still a difficult task which requires a
great computational effort so that the use of kinetic equations still plays
an important role in the solution of problems of practical interest in the field of
rarefied gas dynamics. According to figures (\ref{figA}) and (\ref{figB}), 
there is a good agreement between our results  
and those provided by the other authors when diffuse scattering is assumed. 
The reciprocity relation (\ref{rec7}) is fullfilled within an accuracy
of 0.1\%. Table \ref{tabCA} shows the fullfillment of the reciprocity
relation (\ref{rec7}) when $\delta$=0.1 and 1 for some sets of accommodation
coefficients. 

For other kinds of gas-surface interaction law, some numerical results obtained in the
present work are presented in tables (\ref{tabA}) and (\ref{tabB}) in 
a range of rarefaction parameter $\delta$ which covers the free molecular,
 transitional and hydrodynamic regimes. The values of the accommodation coefficients 
considered in the calculations were chosen because, in practice, the coefficients vary in
the ranges $0.6 \le \alpha_t \le 1$ and $0.1 \le \alpha_n \le 1$ for some
gases, see e.g. \cite{Sha33,Sha113}. According to table \ref{tabA}, the
thermophoretic force can be either in the direction of the temperature gradient
or in the opposite direction to it. Usually, the thermophoretic force is in
the opposite direction to the temperature gradient, i.e. the force tends to move
the particle from hot to cold region. However, in some situations the
movement of the particle from cold to hot region can occur and such a
phenomenon is known as negative thermophoresis, which correponds to a force
in the same direction of the temperature gradient. The negative thermophoresis of particles 
 with high thermal conductivity is expected to appear at large
values of rarefaction parameter, i.e. in the continuum and near-continuum
regimes. Table \ref{tabA} shows the existence of negative thermophoresis when
$\delta$=10 in case of some sets of accommodation coefficients. For instance,
when $\alpha_n$=0.1 and $\alpha_t$ varies from 1 to 0.5, the force is
reversed, i.e. the force is positive in the direction of the temperature gradient. 
Therefore, one can conclude that the occurrence of the negative thermophoresis depends not only
on the rarefaction degree of the gas flow and thermal conductivity of
aerosol particles, but also on the accommodation coefficients on the surface. Therefore, the
appropriate modelling of the gas-surface interaction plays a fundamental role 
for the correct description of the thermophoresis phenomenon. 

According to tables \ref{tabA} and \ref{tabB}, 
the thermophoretic and drag forces depend on both 
accommodation coefficients in the whole range of the gas rarefaction.
As one can see in table \ref{tabA}, for fixed values of $\alpha_n$, when the
thermophoretic force is in the opposite direction to the temperature gradient, 
its magnitude decreases as $\alpha_t$ varies from 1 to 0.5. Moreover,
for fixed values of $\alpha_t$, the magnitude of the thermophoretic force
increases when $\alpha_n$ varies from 1 to 0.1. The same qualitative
behavior is observed in table \ref{tabB} for the drag force on the sphere.
In fact, this kind of behavior is due to the fact that an increase of $\alpha_t$ 
means an increase of tangential stress acting on the sphere, while an increase of $\alpha_n$
means an increase of normal stress on the sphere. However, when the force is reversed, 
the larger $\alpha_t$ and $\alpha_n$ the larger the magnitude of the thermophoretic 
force acting on the sphere.
The results given in these
tables also show us that, in spite of the smallness of the thermophoretic
force in comparison to the drag force on the sphere, the dependence 
of the thermophoretic force on the accommodation
coefficients is larger than that observed for the drag force. 
For instance, the maximum deviation of the
thermophoretic force from the corresponding value for complete accommodation is around 50\%
when $\delta$=0.1, 94\% when $\delta$=1 and larger than 100\% when
$\delta$=10. On the other hand, the maximum deviation of the drag force from
that value in case of diffuse scattering is around 7\% when $\delta$=0.1,
5\% when $\delta$=1 and 0.5\% when $\delta$=10. It can also be seen that the 
dependence of the thermophoretic force 
on the accommodation coefficients is larger in the hydrodynamic regime, 
while for the drag force such a dependence is larger in the free molecular regime.

\subsection{Slip flow regime}

In situations where $\delta >> 1$, the drag force acting on the sphere 
can be obtained from the solution of the Navier-Stokes equations with slip
boundary condition. Thus, according to \cite{Sha84}, the dimensionless drag force is
written as 
\begin{equation}
F_u=\frac {3}{2\delta}
\left(1-\frac{\sigma_{\mbox{\tiny{P}}}}{\delta}\right),
\lae{res1a}
\end{equation}
where $\sigma_{\mbox{\tiny{P}}}$ is the viscous
slip coefficient, which strongly depends on the TMAC. 
\cite{Sha43,Sha44} provide some values of $\sigma_{\mbox{\tiny{P}}}$ for a
single gas obtained numerically via the solution of the Shakhov kinetic
equation and the Cercignani-Lampis model of gas-surface
interaction. For practical
applications, a formula which perfectly
interpolates the results provided by \cite{Sha43,Sha44} is presented by
\cite{Sha84} as follows
\begin{equation}
\sigma_{\mbox{\tiny{P}}}=\frac{1.772}{\alpha_t}-0.754.
\lae{res2a}
\end{equation}   
The results obtained from (\ref{res1a}) are in good agreement with the 
experimental data found in the literature. For instance, in case of $\alpha_t$=1 and $\delta > 10$, the results
obtained from (\ref{res1a}) agree very well with the experimental data provided by 
\cite{Allen,Hutchins} for the drag force on spherical particles of polystyrene 
latex in air at ambient conditions.   

Similarly, for situations where $\delta > 10$, an expression for the the thermophoretic force on
the sphere can be obtained from the solution of the Navier-Stokes-Fourier equations 
with slip velocity and temperature jump boundary conditions, e.g. \cite{Bro06}. However, 
the thermophoretic force predicted from these equations have a large
deviation from experimental data for high thermal conductivity particles, e.g. \cite{Brock1965}. 
This failure is attributed to the fact that the first-order slip solution cannot account for
the phenomena arising in the vicinity of the particle, which is a region with large departure from
local thermodynamic equilibrium. On the contrary, theories based on higher-order approximation in the
Knudsen number, as that proposed by \cite{Son02}, are able to predict the thermophoretic
force on high thermal conductivity particles in the continuum and near
continuum or slip flow regimes. However, is worth mentioning that the
solutions obtained from higher order approximations in the Knudsen number
rely on the accurate determination of the slip
velocity and temperature jump coefficients, and these quantities can be
strongly dependent on the accommodation coefficients and intermolecular interaction potential.
As pointed out in the review by \cite{Sha84}, to obtain more reliable theoretical values
of the slip and jump coefficients, numerical methods to solve the Boltzmann
equation with a realistic potential of intermolecular interaction and 
new models of the gas-surface interaction should be developed as well as more experiments
shoud be carried out.  

\subsection{Flow field}

The macrocroscopic characteristics of the gas flow around the sphere are 
dependent on the accommodation coefficients. Figures \ref{figC}-\ref{figF} show 
the profiles of the radial and polar components of the bulk velocity, density and 
 temperature deviations from equilibrium, respectively, due to the 
thermodynamic force $X_u$, as functions of the distance $r/\delta$, when $\delta$=0.1, 1 and
10. Similarly, Figures \ref{figG}-\ref{figJ}
show the same profiles, but due to the thermodynamic force $X_T$. Note that, according 
to the definitions given in (\ref{a3}) and (\ref{a7}), the dimensionless
distance $r/\delta$ corresponds to the ratio of the dimension radial distance
$r'$ from the sphere to the radius $R_0$ of the sphere. 
The dependence on the TMAC is 
shown in Figures \ref{figC}, \ref{figD}, \ref{figG} and
\ref{figH} with the NEAC fixed at
$\alpha_n$=0.1. The dependence on the NEAC is 
shown in Figures \ref{figE}, \ref{figF}, \ref{figI} and \ref{figJ} with 
the TMAC fixed at $\alpha_t$=1.
The fixed values of $\alpha_n$ and 
$\alpha_t$ were chosen because, as can be noted in tables \ref{tabA} and
\ref{tabB}, the larger deviations of the thermophoretic and drag forces 
from those values in case of diffuse scattering occurs when
$\alpha_n$=0.1 and $\alpha_t$=1. 
From figures \ref{figC}-\ref{figF}, regarding the macroscopic quantities due 
to the thermodynamic force $X_u$, one can conclude the following.
(i) According to figures \ref{figC} and \ref{figE}, near the sphere 
the radial component of the bulk velocity tends to decrease while the
the polar component tends to increase. This situation corresponds to a 
decrease in the the bulk velocity of the gas flow due to the presence of the
sphere. This qualitative behavior is already known
 from the literature. However, as one can see from
these figures, although the bulk velocity of the gas flow does not depend
on the NEAC $\alpha_n$, theres is a
small dependence on the TMAC $\alpha_t$. (ii) Regarding the number density and 
temperature of the gas
flow around the sphere, according to figures \ref{figD} and \ref{figF},
these quantities always decrease in the vicinity of the sphere. However,
while the dependence of the density and temperature deviations
 on the TMAC is negligible, there is
a significant dependence on the NEAC. As one can see in figure \ref{figF}, when
$\alpha_n$ varies from 1 to 0.1, the temperature deviation near the sphere strongly deviates
from the corresponding plot for diffuse scattering in the three
situations of gas rarefaction considered, i.e. $\delta$=0.1, 1 and 10. The dependence
of the density deviation on the NEAC is 
negligible in the free molecular regime, but it tends to be significant as the gas
flow tends to the hydrodynamic regime. 

From figures \ref{figG}-\ref{figJ}, regarding the macroscopic quantities due
to the thermodynamic force $X_T$, one can conclude the following. (i)
According to figures \ref{figG} and \ref{figI}, the radial and polar
components of the bulk velocity depends on both accommodation coefficients.
However, the dependence on the TMAC tends to be larger as the gas flow tends to
 the free molecular
regime, while the dependence on the NEAC tends to be larger as the gas flow tends to 
the hydrodynamic regime. Quantitatively, the dependence on the NEAC
 is larger than that on the TMAC. Figure \ref{figI} shows us that when $\delta$=10,
$\alpha_n$=$\alpha_t$=1, the bulk velocity of the gas in the vicinity of the sphere starts to 
change direction, which means the starting of the negative thermophoresis due to
the thermal creep around the sphere. In order to see the dependence of the bulk
velocity of the gas flow due to the temperature gradient on the NEAC
 as well as the appearance of negative thermophoresis, the the radial and polar components of 
the bulk velocity as functions of the distance $r/\delta$ are given
in figure \ref{figK} in case of $\delta$=10, $\alpha_n$ fixed at 0.1 and
$\alpha_t$ varying from 1 to 0.5. According to this figure, the bulk velocity of the gas
flow depends strongly on the NEAC and is even reversed when $\alpha_t < 1$, which means a gas flow in the opposite
direction to the temperature gradient, i.e. the negative thermophoresis.   
 (ii) The temperature of the gas flow tends to increase in the vicinity of the sphere
in free molecular and transition regimes. However, in the hydrodynamic
regime the temperature of the gas decreases near the sphere.
The qualitative behavior is the same for arbitrary values of accommodation coefficients,
 but the influence of the NEAC is larger than that on 
the TMAC.

\subsection{Comparison with experiment}

Figure \ref{figexp} shows the comparison between the results obtained in the
present work for the dimension thermophoretic force, in $\mu$N, and those provided by \cite{Bos02}
for a copper sphere in argon gas in a wide range of the gas rarefaction.
 The experimental apparatus employed by \cite{Bos02} consisted on measuring the thermophoretic force acting
on a sphere, with radius of about 0.025 m, and fixed in the middle of two 61 cm x
61 cm x 0.9 cm copper plates placed within a vacuum chamber filled with argon gas at ambient
temperature and pressure ranging from 13.3 Pa to 0.013 Pa. The distance
betwen the plates was fixed at 40 cm and their temperatures were set up to establish a temperature 
gradient of 35 K/m between them. However, in rarefied conditions, the
profile of the gas temperature between the plates is not linear and there is
a temperature jump at the plates. Moreover, the temperature gradient through the gas tends 
to decrease as the pressure decreases. Therefore, to compare our 
numerical results with the experimental data provided by \cite{Bos02}, these effects must be
taken into account. Thus, for pressures lower than 1 Pa, the temperature gradient was estimated with 
basis on the temperature profiles presented by \cite{Bos02} in figure 5, 
which were obtained numerically via the DSMC method. The temperature jump
coefficient necessary to estimate the gas temperature at the walls was
obtained from the interpolating formula given by \cite{Sha113}. The numerical 
calculations were carried out for $\alpha_t$=1 and $\alpha_n$=0.9, values
recommended by \cite{Sha113} for argon gas at 
ambient temperature and metallic surface.
 The viscosity of the gas was obtained
from \cite{Vog01}. Thus, according to figure \ref{figexp}, there is
a good agreement between the results obtained in the present work and the
experimental data provided by \cite{Bos02}. However, although the negative
thermophoresis was detected in the experiment carried out by \cite{Bos02},
in the present work it was not predicted numerically for the chosen set of
the accommodation coefficients. Nonetheless, according to the results presented in table \ref{tabA}, 
the negative thermophoresis is predicted numerically for some sets of
accommodation coefficients when $\delta$=10.

\section{Concluding remarks}

In the present work, the classical problems of thermophoresis and viscous
drag on a sphere with high thermal conductivity were investigated on the basis of
the linearized kinetic equation proposed by Shakhov and the Cercignani-Lampis model
 to the gas-surface interaction law. In the free molecular regime the 
solutions for both problems were obtained analytically, while in the
transitional and hydrodynamic
regimes the problems were solved numerically via
the discrete velocity method with a proper method to take into account the
discontinuity of the distribution function around a convex body. The
reciprocity relation between the cross phenomena was obtained and verified
numerically within an accuracy of 0.1\%. 
The thermophoretic and drag forces acting on the sphere, 
as well as the macroscopic characteristics of the gas flow
around it, were obtained for some sets of TMAC and NEAC obtained from
experiments for monoatomic gases at various surfaces. The results show a strong dependence of the
thermophoretic force on both accommodation coefficients, including the
appearance of the negative thermophoresis in the hydrodynamic regime as TMAC
and the NEAC vary. Moreover, the results show a dependence 
of the viscous drag force on the accommodation  coefficients, but such a dependence 
is smaller than that observed for the thermophoretic force. Similarly, the
flow fields around the sphere also depend on 
the accommodation coefficients, but the dependence of the quantities due to the
 thermodynamic force $X_u$ is smaller than that observed for the quantities due 
to the thermodynamic force $X_T$. As the gas tends to the hydrodynamic
regime, the drag force tends to be independent of the NEAC as predicted by
the expression obtained from the Navier-Stokes equations with slip
conditions. Regarding the comparison with experimental data, a good agreement
was verified for the case of a copper sphere in argon gas and the negative thermophoresis
was predicted as dependent of the proper choice of the TMAC and NEAC.  
According to the results presented in the
present work, a better understanding of the transport
phenomena of spherical aerosols relies on the correct description of the
gas-surface interaction law so that further investigations on this research topic
should be carried out, experimentally and numerically. Moreover, since the data on
this topic are still scarce, the results provided in the present work represent a 
significant contribution towards a better understanding of the phoretic
phenomena.  


\section*{Acknowledgments}

The present calculations were carried out at the Laborat\'orio Central de
Processamento de Alto Desempenho (LCPAD) of Universidade Federal do Paran\'a
(UFPR, Brazil). The authors acknowledge the Conselho Nacional de
Desenvolvimento Cient\'{\i}fico e Tecnol\'ogico (CNPq), grant 304831/2018-2, and the
Funda\c{c}\~ao de Amparo \`a Pesquisa do Estado de S\~ao Paulo (FAPESP), grant
2015/20650-5, for the support of the research.


\bibliographystyle{unsrt}

\clearpage




\begin{table}
\centering
\begin{threeparttable}
\begin{tabular}{cccccccc}
 & & \multicolumn{6}{c}{$F_{\mbox{\tiny{T}}}$}\\ \cline{3-8}
$\alpha_t$ & $\alpha_n$ &\multicolumn{2}{c}{$\delta\rightarrow 0$} & 
 \multicolumn{2}{c}{$\delta$=0.1} & \multicolumn{2}{c}{$\delta$=1}  \\ \hline
 & & Eq. (\ref{rec7}) & Eq. (\ref{fm10})\tnote{a} &
 Eq. (\ref{rec7}) & Eq. (\ref{ke9a})\tnote{b} & Eq. (\ref{rec7}) & Eq.
(\ref{ke9a})\tnote{b} \\ \cline{3-8}
1  & 1  &-0.2821 &-0.2821 &-0.2725 &-0.2725 &-0.1731 &-0.1729 \\
   & 0.7&-0.3299 &-0.3299 &-0.3221 &-0.3221 &-0.2280 &-0.2279 \\
   & 0.5&-0.3596 &-0.3596 &-0.3530 &-0.3530 &-0.2643 &-0.2642  \\
   & 0.1&-0.4119 &-0.4119 &-0.4093 &-0.4093 &-0.3358 &-0.3357 \\[0.25cm]
0.5& 1  &-0.2109 &-0.2109 &-0.2036 &-0.2035 &-0.1185 &-0.1184 \\
   & 0.7&-0.2594 &-0.2594 &-0.2531 &-0.2531 &-0.1754 &-0.1753 \\
   & 0.5&-0.2887 &-0.2887 &-0.2840 &-0.2840 &-0.2129 &-0.2128 \\
   & 0.1&-0.3414 &-0.3414 &-0.3404 &-0.3404 &-0.2866 &-0.2865 \\[0.25cm]
0.1& 1  &-0.1551 &-0.1551 &-0.1468 &-0.1469 &-0.0648 &-0.0649  \\
   & 0.7&-0.2030 &-0.2030 &-0.1964 &-0.1964 &-0.1221 &-0.1223 \\
   & 0.5&-0.2326 &-0.2326 &-0.2273 &-0.2273 &-0.1603 &-0.1601\\
   & 0.1&-0.2872 &-0.2880 &-0.2855 &-0.2862 &-0.2368 &-0.2350 \\ \hline
\end{tabular}
\begin{tablenotes}
\footnotesize
\item[a] Analytical solution in the free molecular regime. 
\item[b] Numerical solution due to $X_T$ ($n$=$T$).
\end{tablenotes}
\end{threeparttable}
\caption{Thermophoretic force on the sphere: verification of the reciprocity relation.}
\lae{tabCA}
\end{table}

\clearpage

\begin{table}
\centering
\begin{threeparttable}
\begin{tabular}{ccccccc} 
 & & \multicolumn{4}{c}{$F_{\mbox{\tiny{T}}}$}\\ \cline{3-7}
             & $\alpha_t$ & $\alpha_n$=0.1 & 0.5 & 0.8  & 0.9  & 1.0\\ \hline
$\delta \rightarrow 0$\tnote{a}  & 0.5 &-0.3414 &-0.2887 &-0.2434 &-0.2273 &-0.2109 \\
             & 0.8 &-0.3837 &-0.3314 &-0.2863 &-0.2703 &-0.2539 \\
             & 0.9 &-0.3978 &-0.3455 &-0.3004 &-0.2844 &-0.2680 \\
             & 1.0 &-0.4119 &-0.3596 &-0.3145 &-0.2985 &-0.2821 \\[0.25cm]
$\delta$=0.01& 0.5 &-0.3419 &-0.2887 &-0.2434 &-0.2273 &-0.2109 \\
             & 0.8 &-0.3840 &-0.3309 &-0.2856 &-0.2696 &-0.2531 \\
             & 0.9 &-0.3981 &-0.3450 &-0.2996 &-0.2836 &-0.2672 \\
             & 1.0 &-0.4122 &-0.3590 &-0.3137 &-0.2977 &-0.2812 \\[0.25cm]
$\delta$=0.1 & 0.5 &-0.3404 &-0.2840 &-0.2365 &-0.2200 &-0.2031 \\
             & 0.8 &-0.3817 &-0.3255 &-0.2785 &-0.2613 &-0.2444\\
             & 0.9 &-0.3955 &-0.3393 &-0.2922 &-0.2757 &-0.2581\\
             & 1.0 &-0.4093 &-0.3530 &-0.3059 &-0.2894 & -0.2725\\[0.25cm]
$\delta$=1   & 0.5 &-0.2865 &-0.2129 &-0.1502 &-0.1320 &-0.1136 \\
             & 0.8 &-0.3165 &-0.2445 &-0.1894 &-0.1633 &-0.1455\\
             & 0.9 &-0.3263 &-0.2546 &-0.1997 &-0.1813 &-0.1554\\
             & 1.0 &-0.3360 &-0.2645 &-0.2099 & -0.1915 &-0.1730 \\[0.25cm]

$\delta$=10  & 0.5 &-0.0643 &-0.0191 & 0.00663& 0.01379 & 0.02054 \\
             & 0.8 &-0.0661 &-0.0296 &-0.00769&-0.00146 & 0.00407\\
             & 0.9 &-0.0668 &-0.0319 &-0.0109 &-0.00473 & 0.00072 \\
             & 1.0 &-0.0674 &-0.0339 &-0.0136 &-0.00765 &-0.00199 \\ \hline
\end{tabular}
\begin{tablenotes}
\footnotesize
\item[a] Eq. (\ref{fm10}), free molecular regime.
\end{tablenotes}
\end{threeparttable}
\caption{Dimensionless thermophoretic force on the sphere.}
\label{tabA}
\end{table}

\clearpage
\begin{table}
\centering
\begin{threeparttable}
\begin{tabular}{ccccccc}
 & & \multicolumn{4}{c}{$F_u$}\\ \cline{3-7}
             & $\alpha_t$ & $\alpha_n$=0.1 & 0.5 & 0.8 & 0.9 & 1.0  \\ \hline
$\delta\rightarrow 0$\tnote{a}  & 0.5 & 0.9312 & 0.8979 & 0.8746 & 0.8670 & 0.8596 \\
             & 0.8 & 1.0441 & 1.0107 & 0.9874 & 0.9799 & 0.9724 \\
             & 0.9 & 1.0817 & 1.0483 & 1.0250 & 1.0175 & 1.0100 \\
             & 1.0 & 1.1193 & 1.0859 & 1.0626 & 1.0551 & 1.0477 \\[0.25cm]
$\delta$=0.01& 0.5 & 0.9276 & 0.8940 & 0.8708 & 0.8633 & 0.8559 \\
             & 0.8 & 1.0397 & 1.0062 & 0.9830 & 0.9755 & 0.9681 \\
             & 0.9 & 1.0771 & 1.0436 & 1.0204 & 1.0129 & 1.0055 \\
             & 1.0 & 1.1144 & 1.0810 & 1.0578 & 1.0503 & 1.0429 \\[0.25cm]
$\delta$=0.1 & 0.5 & 0.8954 & 0.8631 & 0.8388 & 0.8316 & 0.8245 \\
             & 0.8 & 1.0027 & 0.9707 & 0.9484 & 0.9387 & 0.9316 \\
             & 0.9 & 1.0382 & 1.0063 & 0.9841 & 0.9769 & 0.9671 \\
             & 1.0 & 1.0738 & 1.0419 & 1.0197 & 1.0125 & 1.0054 \\[0.25cm]
$\delta$=1   & 0.5 & 0.6431 & 0.6237 & 0.5836 & 0.5796 & 0.5757 \\
             & 0.8 & 0.7137 & 0.6953 & 0.6824 & 0.6456 & 0.6419 \\
             & 0.9 & 0.7361 & 0.7179 & 0.7052 & 0.7011 & 0.6627\\
             & 1.0 & 0.7579 & 0.7400 & 0.7275 & 0.7234 & 0.7194 \\[0.25cm] 
$\delta$=10  & 0.5 & 0.1318 & 0.1309 & 0.1306 & 0.1305 & 0.1304 \\
             & 0.8 & 0.1397 & 0.1393 & 0.1391 & 0.1391 & 0.1390 \\
             & 0.9 & 0.1403 & 0.1401 & 0.1400 & 0.1399 & 0.1398 \\
             & 1.0 & 0.1436 & 0.1431 & 0.1430 & 0.1430 & 0.1429 \\ \hline
\end{tabular}
\begin{tablenotes}
\footnotesize
\item[a] Eq. (\ref{fm9}), free molecular regime.
\end{tablenotes}
\end{threeparttable}
\caption{Dimensionless viscous drag force on the sphere}
\label{tabB}
\end{table}

\clearpage


\begin{figure}
\centering
\includegraphics[scale=1]{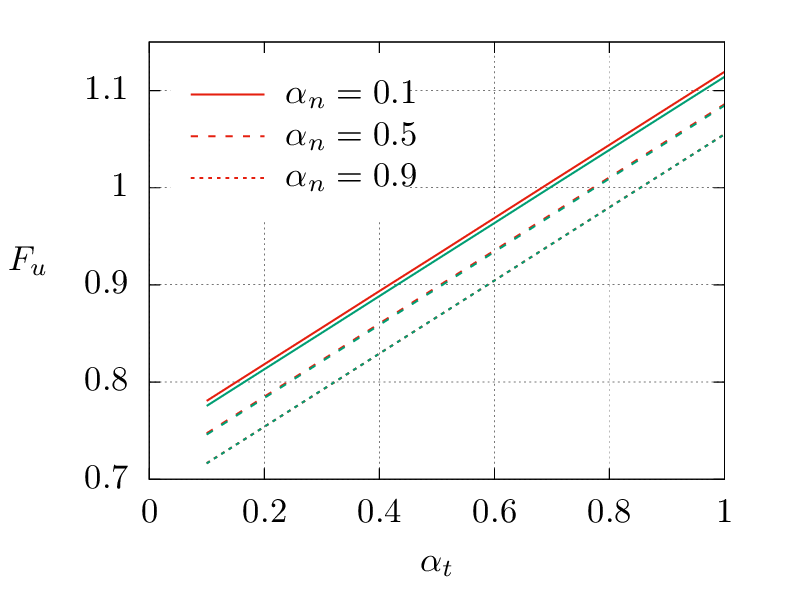}
\caption{Viscous drag force on the sphere in the free molecular regime as function of the
tangential momentum accommodation coefficient. Green lines correspond to the results
presented by \cite{Che25}. Red lines correspond to the present results.}
\lae{figCher1}
\end{figure}

\begin{figure}
\centering
\includegraphics[scale=1]{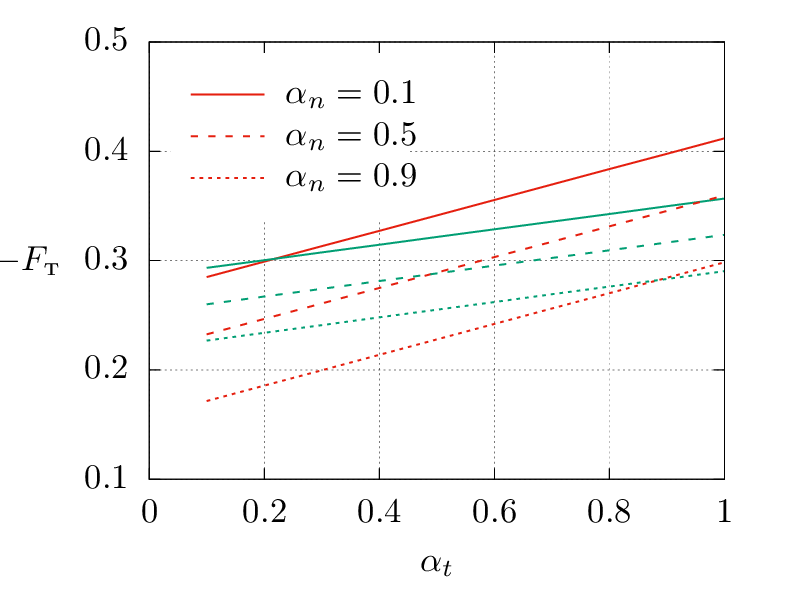}
\caption{Thermophoretic force on the sphere in the free molecular regime as function of the
tangential momentum accommodation coefficient. Green lines correspond to the results
presented by \cite{Che25}. Red lines correspond to the present results.}
\lae{figCher2}
\end{figure}

\begin{figure}
\centering
\includegraphics[scale=1.]{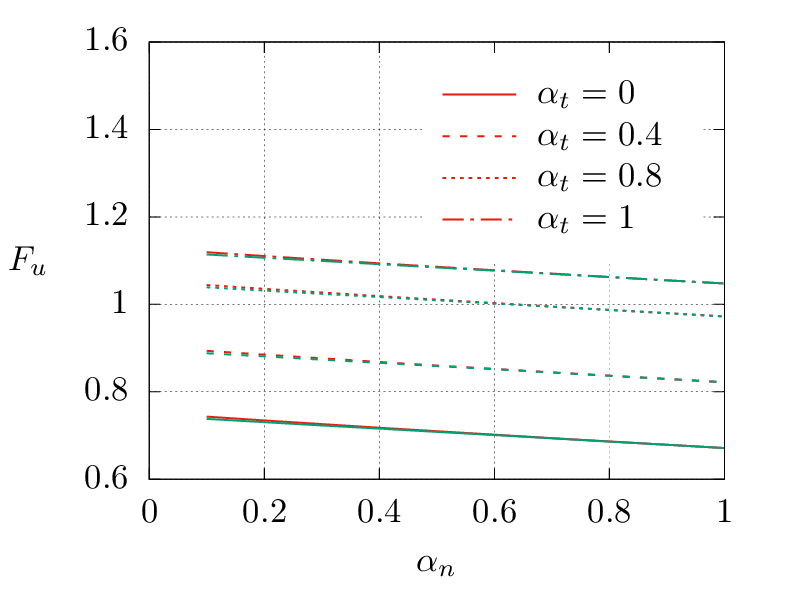}
\caption{Viscous drag force on the sphere in the free molecular regime as function of the
normal energy accommodation coefficient. Green lines correspond to the results
presented by \cite{Che25}. Red lines correspond to the present results.}
\lae{figCher3}
\end{figure}

\begin{figure}
\centering
\includegraphics[scale=1.]{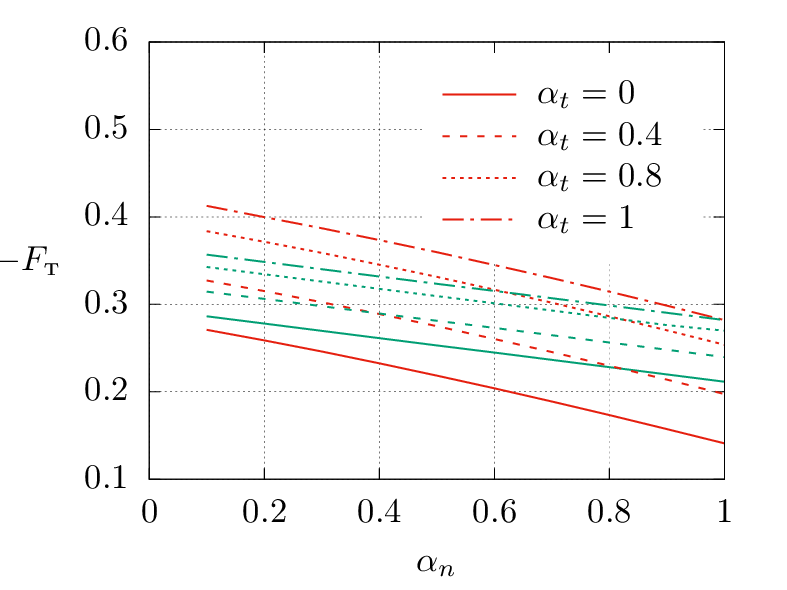}
\caption{Thermophoretic force on the sphere in the free molecular regime as function of the
normal energy accommodation coefficient. Green lines correspond to the results
presented by \cite{Che25}. Red lines correspond to the present results.}
\lae{figCher4}
\end{figure}

\clearpage

\begin{figure}
\centering
\includegraphics[scale=1.]{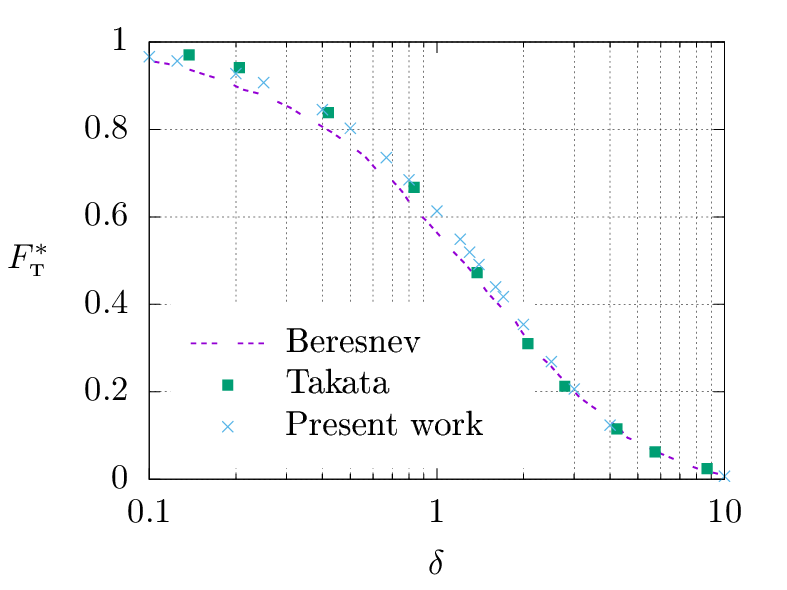}
\caption{Ratio of the thermophoretic force on the sphere to its value in the free molecular
regime: comparison to the results presented by \cite{Ber09,Tak07} for 
diffuse scattering.}
\lae{figA}
\end{figure}

\begin{figure}
\centering
\includegraphics[scale=1.]{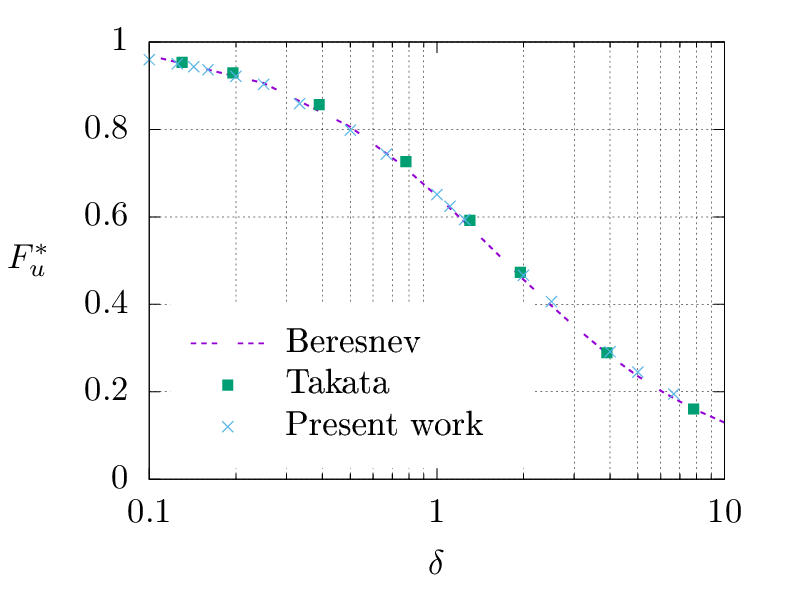}
\caption{Ratio of the drag force on the sphere to its value in the free molecular
regime: comparison to the results presented by \cite{Ber10,Tak09} for diffuse scattering.}
\lae{figB}
\end{figure}

\clearpage

\begin{figure}
\centering
\subfigure[Rarefaction parameter $\delta$=0.1]{
\includegraphics[scale=0.8]{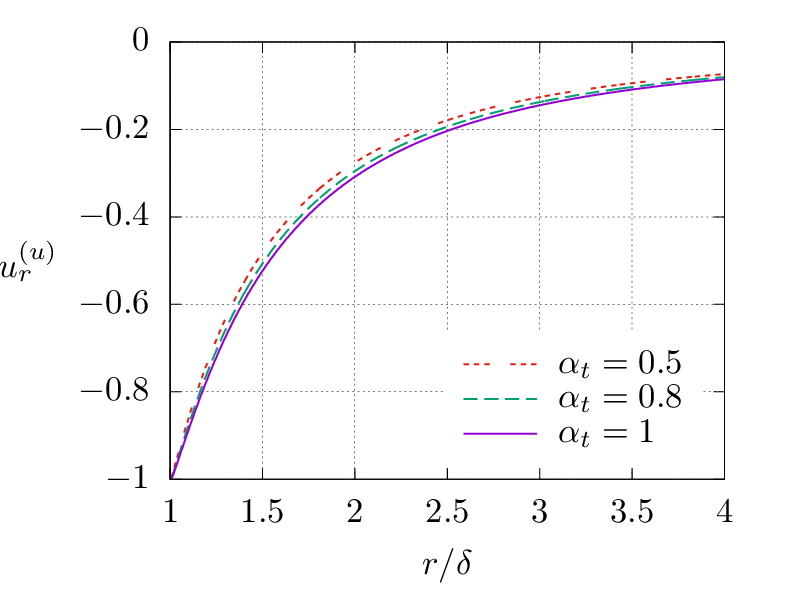}
\includegraphics[scale=0.8]{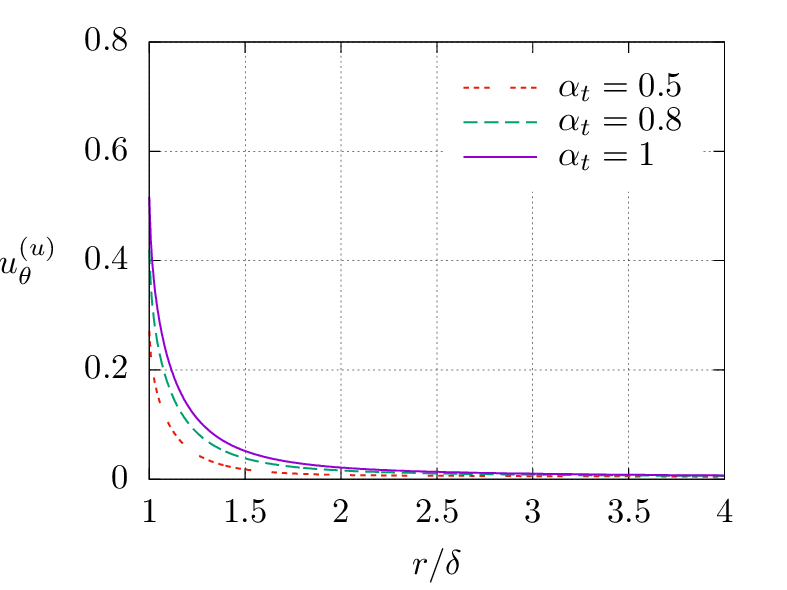}
}
\subfigure[Rarefaction parameter $\delta$=1]{
\includegraphics[scale=0.8]{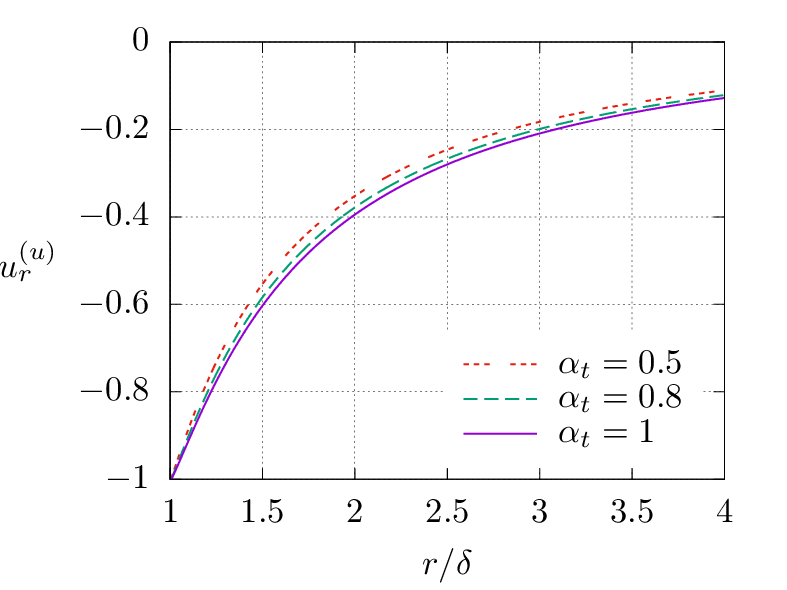}
\includegraphics[scale=0.8]{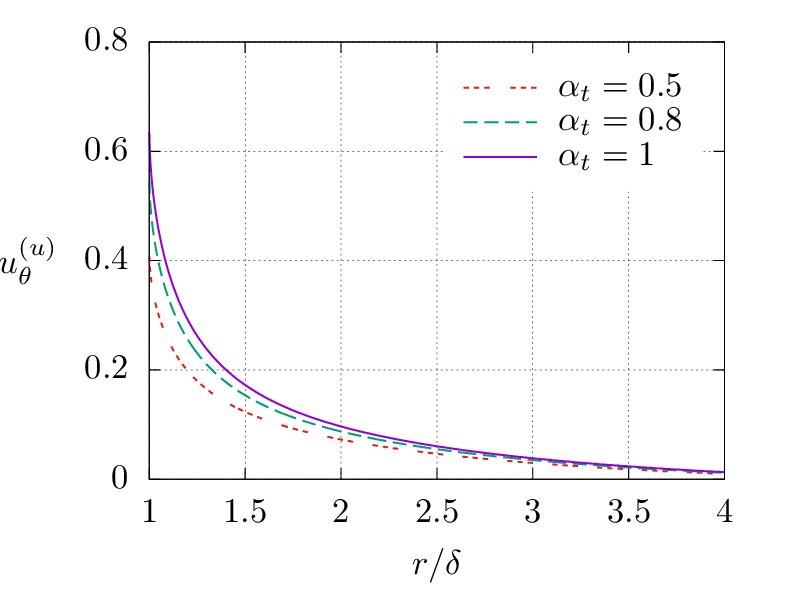}
}
\subfigure[Rarefaction parameter $\delta$=10]{
\includegraphics[scale=0.8]{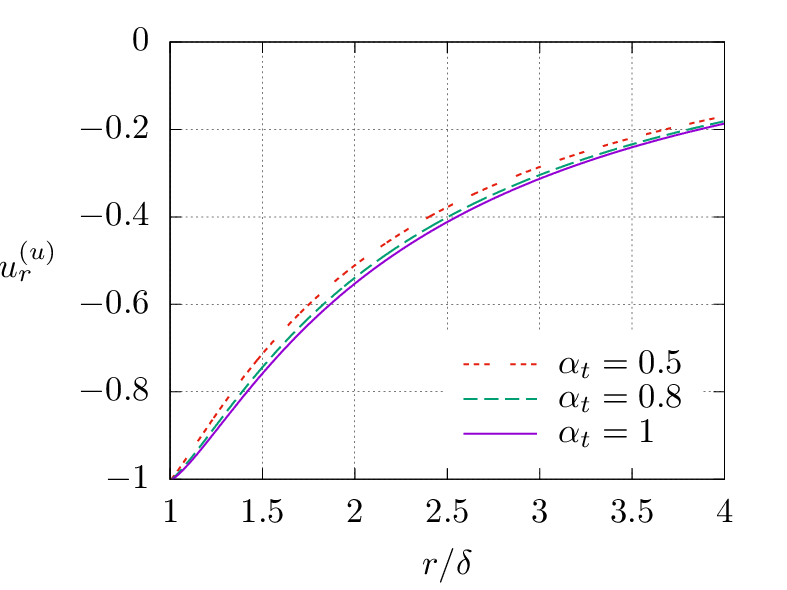}
\includegraphics[scale=0.8]{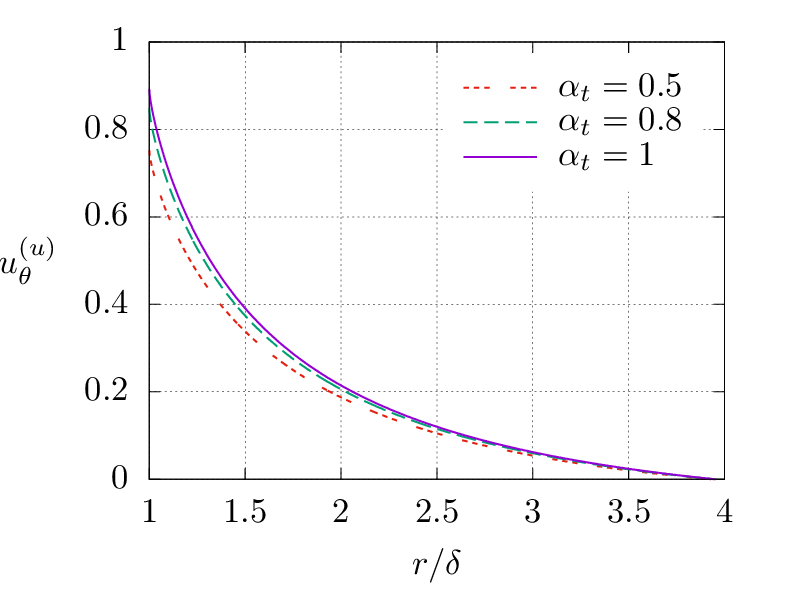}
}
\caption{Components of the bulk velocity as functions of the radial distance from the sphere due to the
thermodynamic force $X_u$ for fixed $\alpha_n$=0.1.}
\lae{figC}
\end{figure}


\begin{figure}
\centering
\subfigure[Rarefaction parameter $\delta$=0.1]{
\includegraphics[scale=0.8]{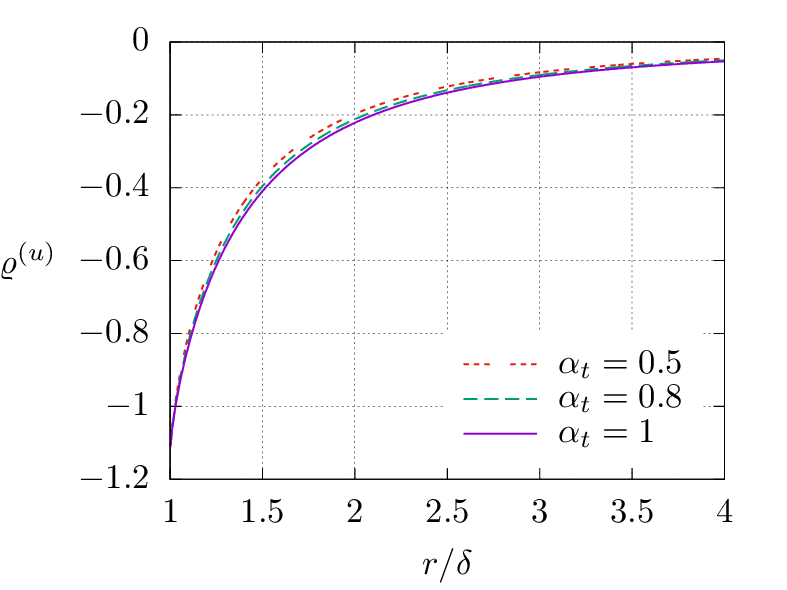}
\includegraphics[scale=0.8]{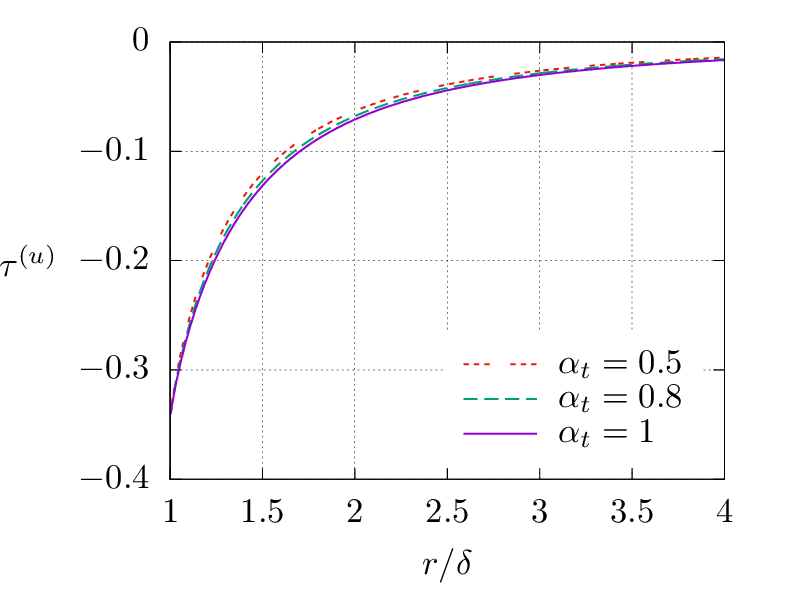}
}
\subfigure[Rarefaction parameter $\delta$=1]{
\includegraphics[scale=0.8]{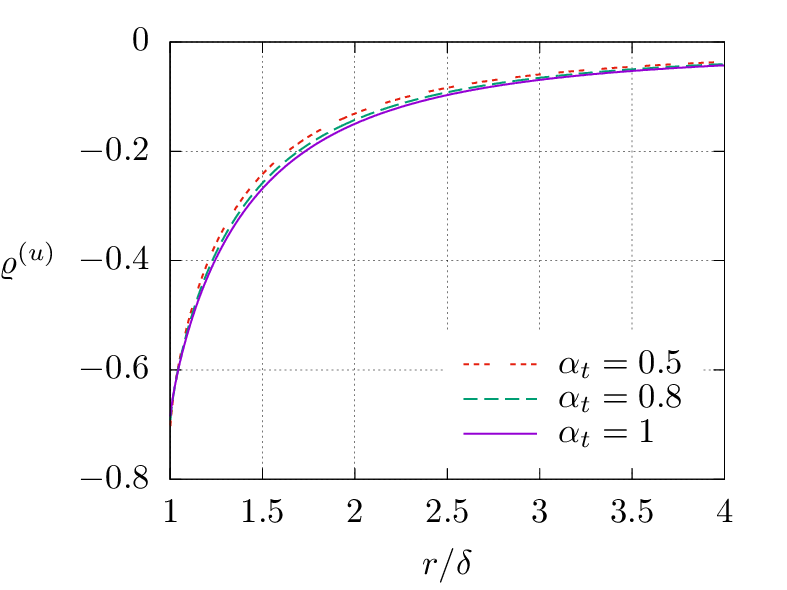}
\includegraphics[scale=0.8]{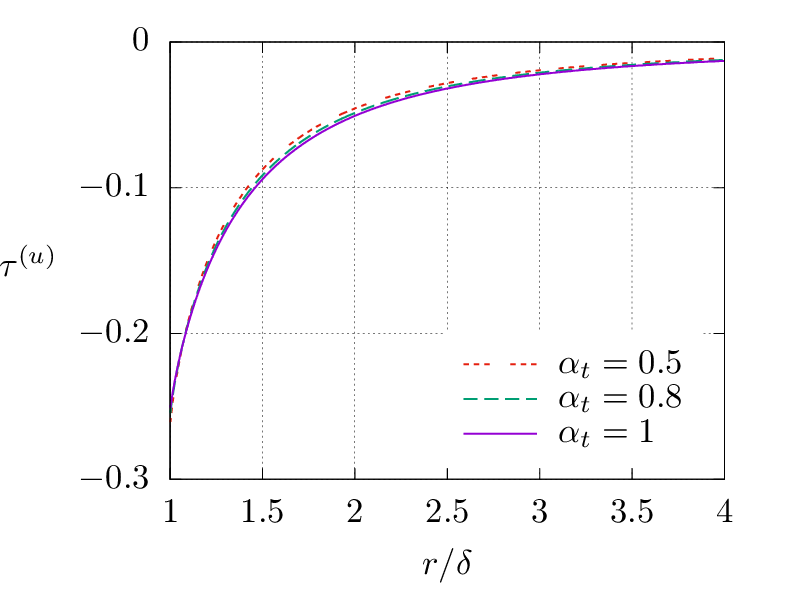}
}
\subfigure[Rarefaction parameter $\delta$=10]{
\includegraphics[scale=0.8]{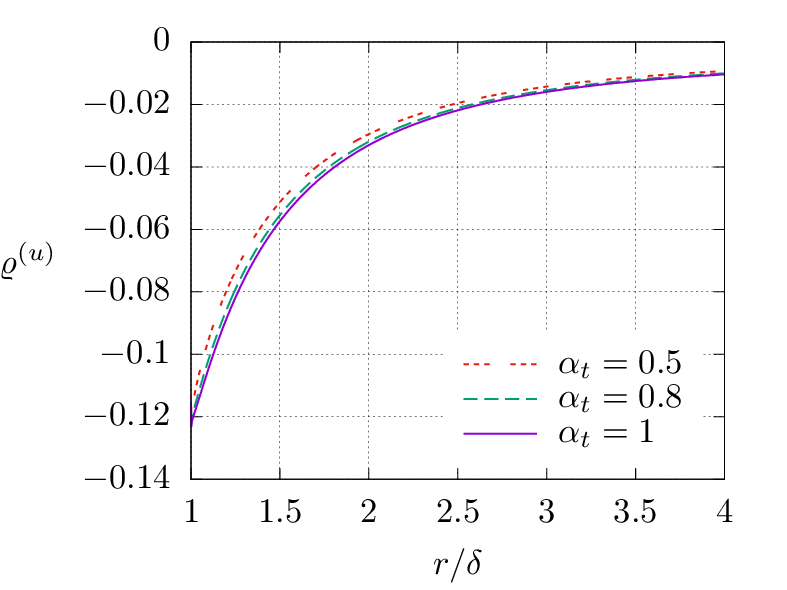}
\includegraphics[scale=0.8]{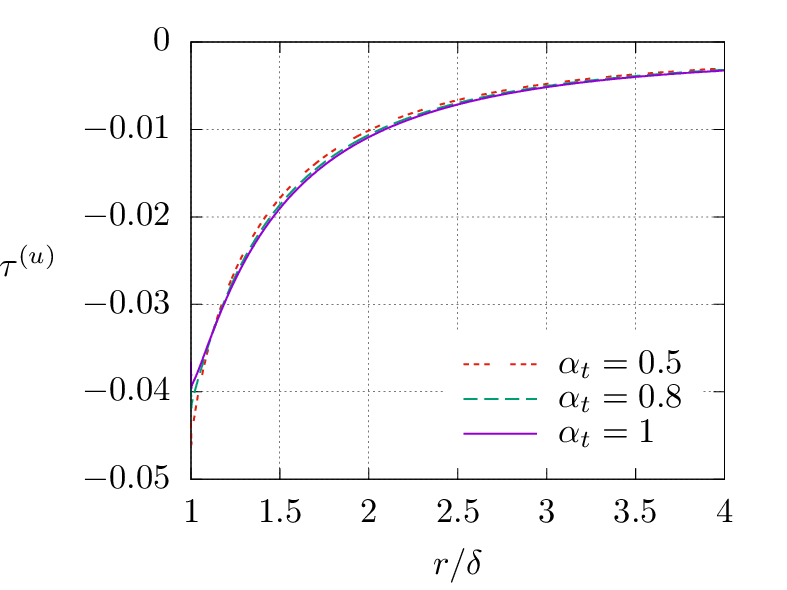}
}
\caption{Density and temperature deviations as functions of the radial
distance from the sphere due to the 
thermodynamic force $X_u$ for fixed $\alpha_n$=0.1.}
\lae{figD}
\end{figure}


\begin{figure}
\centering
\subfigure[Rarefaction parameter $\delta$=0.1]{
\includegraphics[scale=0.8]{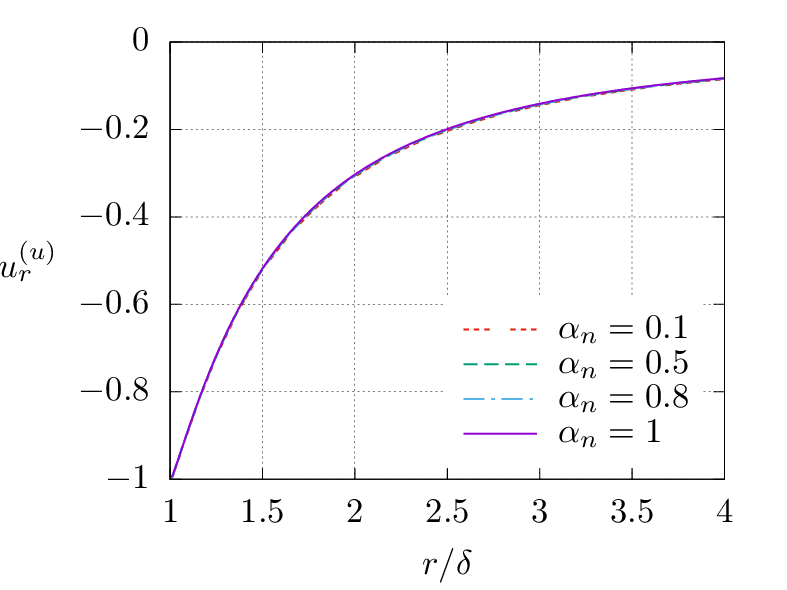}
\includegraphics[scale=0.8]{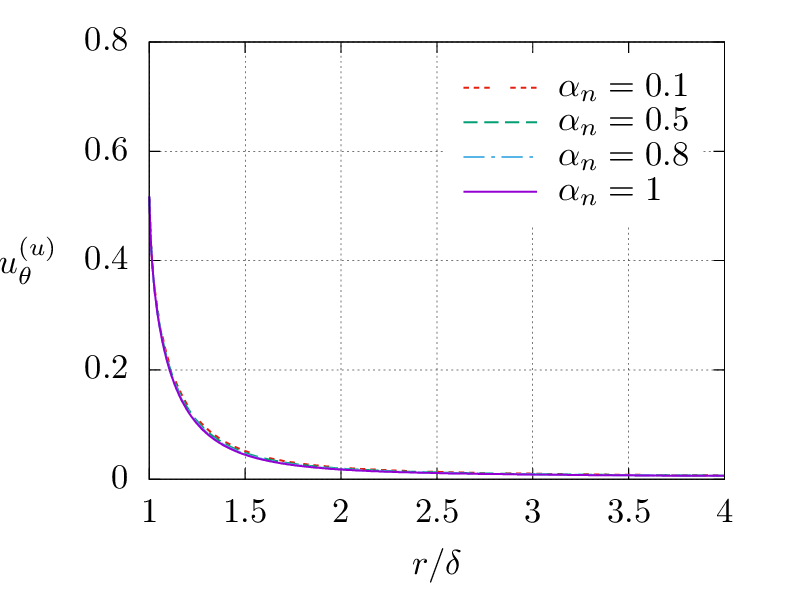}
}
\subfigure[Rarefaction parameter $\delta$=1]{
\includegraphics[scale=0.8]{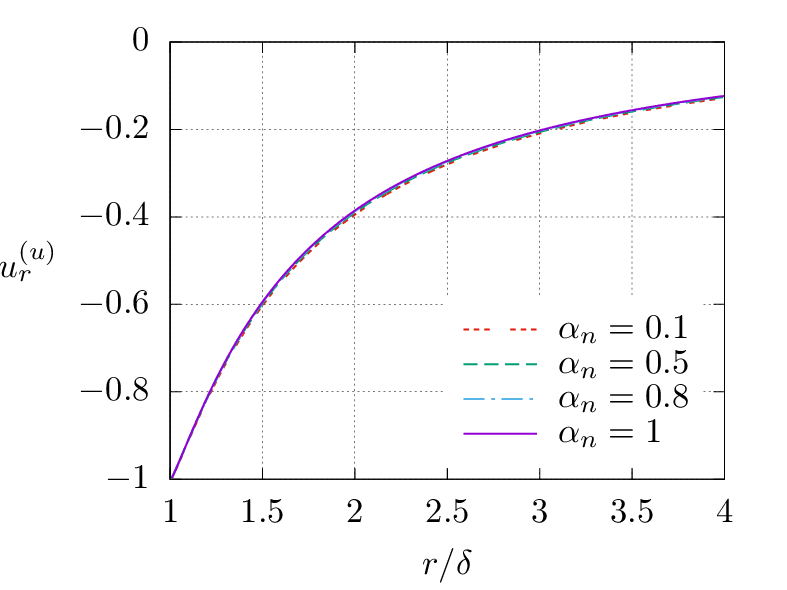}
\includegraphics[scale=0.8]{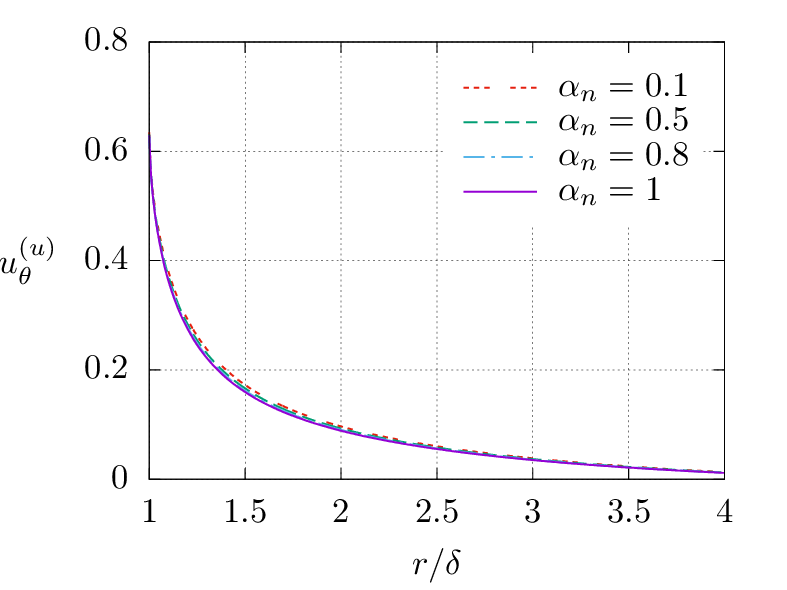}
}
\subfigure[Rarefaction parameter $\delta$=10]{
\includegraphics[scale=0.8]{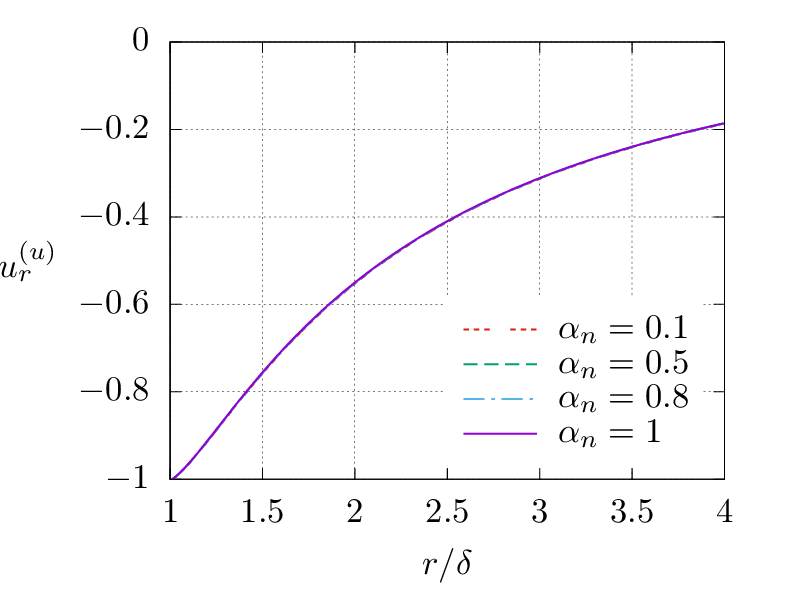}
\includegraphics[scale=0.8]{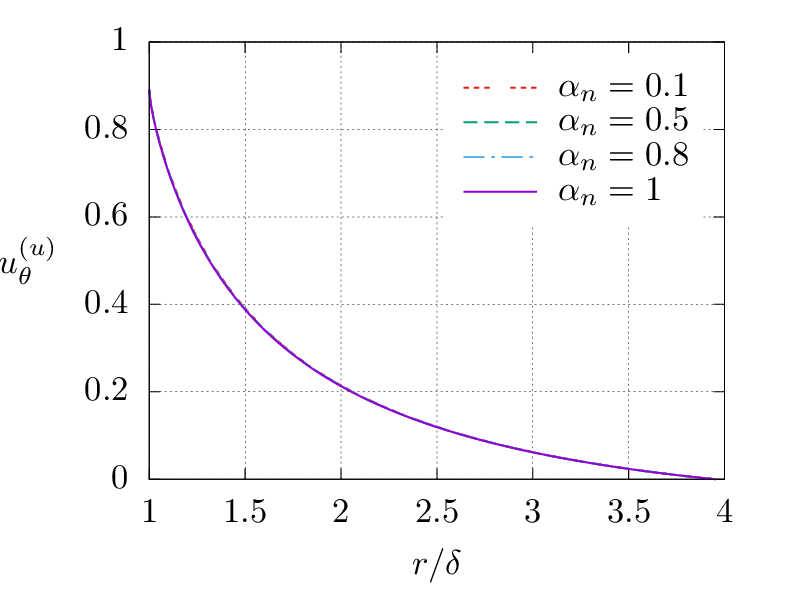}
}
\caption{Components of the bulk velocity as functions of the radial distance
from the sphere due to the
thermodynamic force $X_u$ for fixed $\alpha_t$=1.}
\lae{figE}
\end{figure}


\begin{figure}
\centering
\subfigure[Rarefaction parameter $\delta$=0.1]{
\includegraphics[scale=0.8]{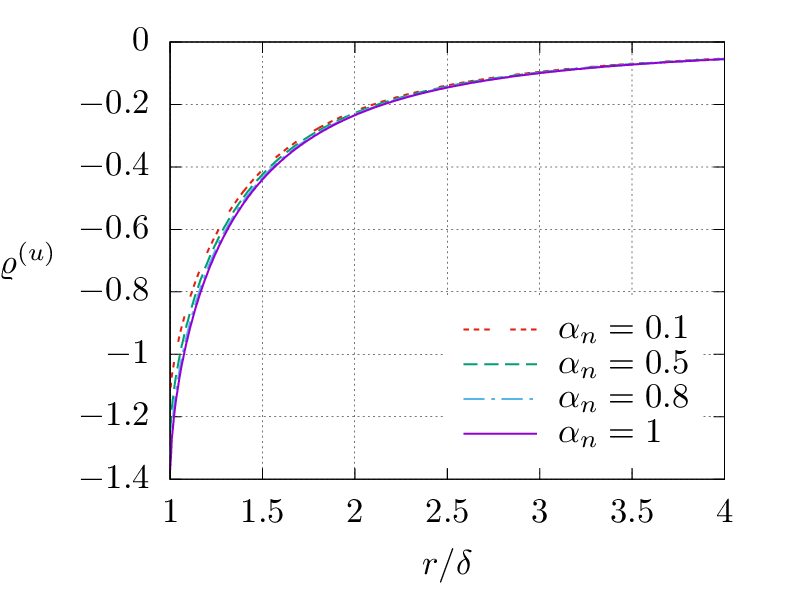}
\includegraphics[scale=0.8]{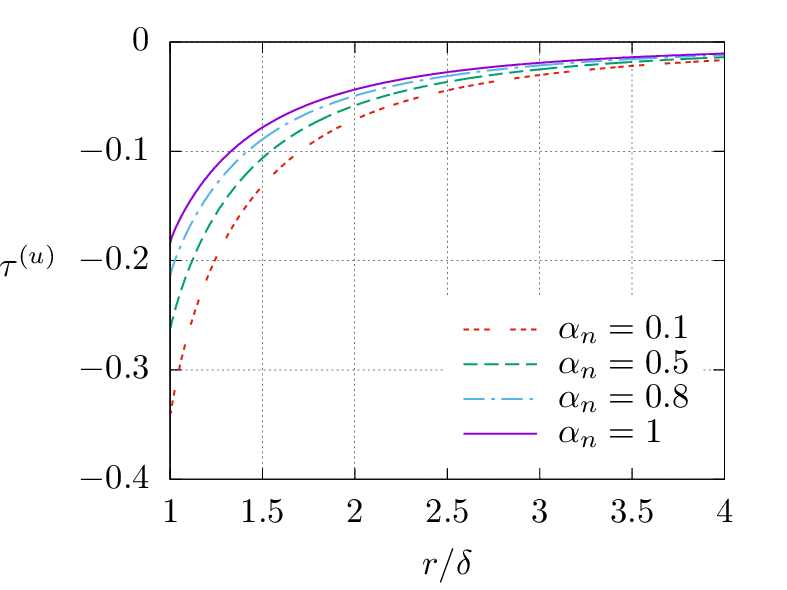}
}
\subfigure[Rarefaction parameter $\delta$=1]{
\includegraphics[scale=0.8]{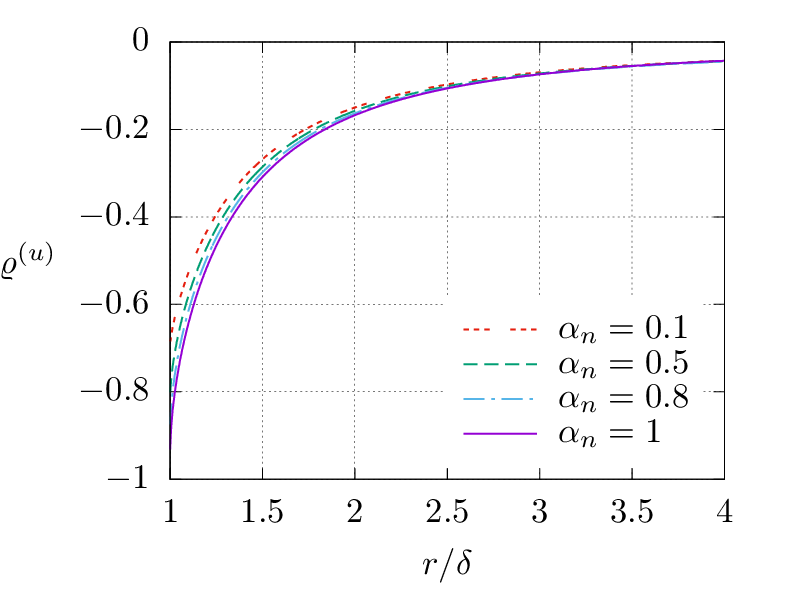}
\includegraphics[scale=0.8]{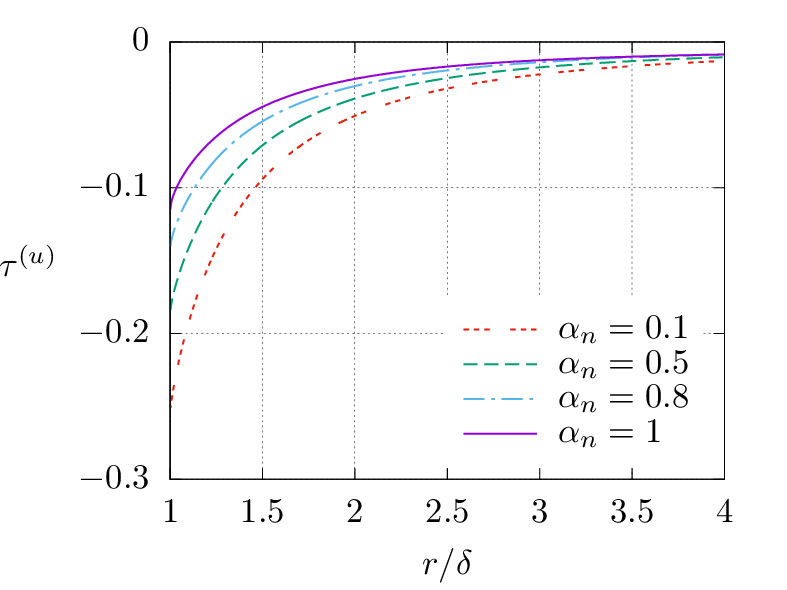}
}
\subfigure[Rarefaction parameter $\delta$=10]{
\includegraphics[scale=0.8]{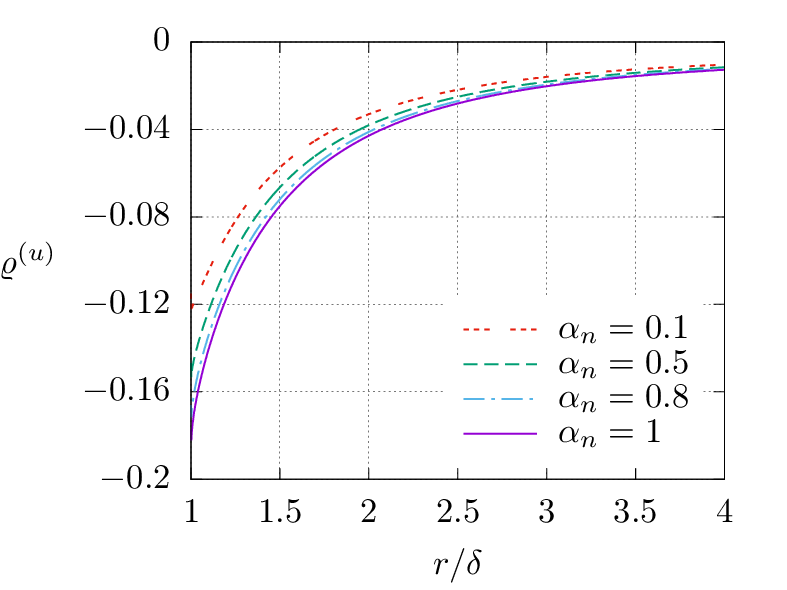}
\includegraphics[scale=0.8]{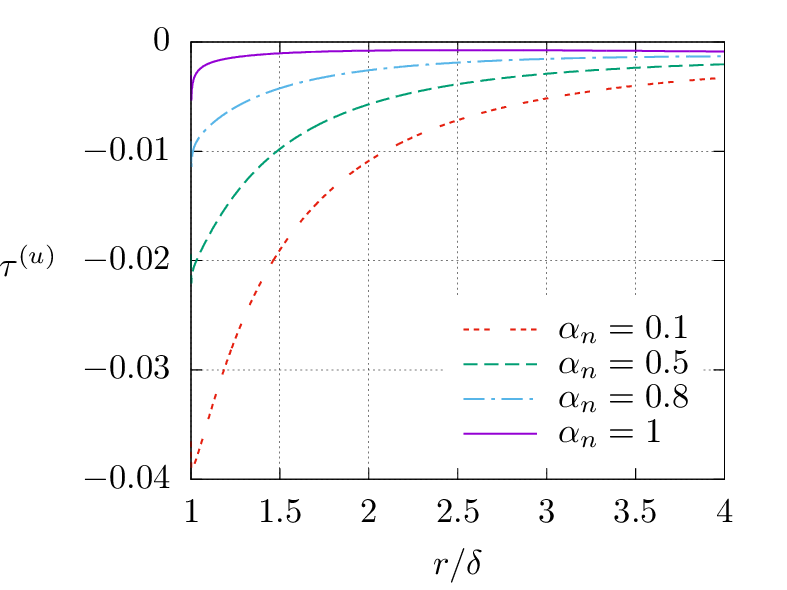}
}
\caption{Density and temperature deviations as functions of the radial
distance from the sphere due to the
thermodynamic force $X_u$ for fixed $\alpha_t$=1.}
\lae{figF}
\end{figure}


\begin{figure}
\centering
\subfigure[Rarefaction parameter $\delta$=0.1]{
\includegraphics[scale=0.8]{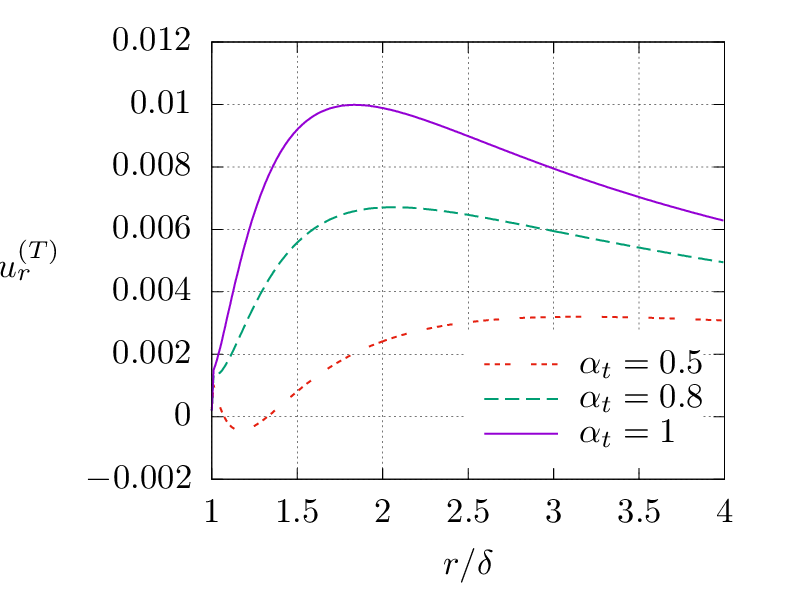}
\includegraphics[scale=0.8]{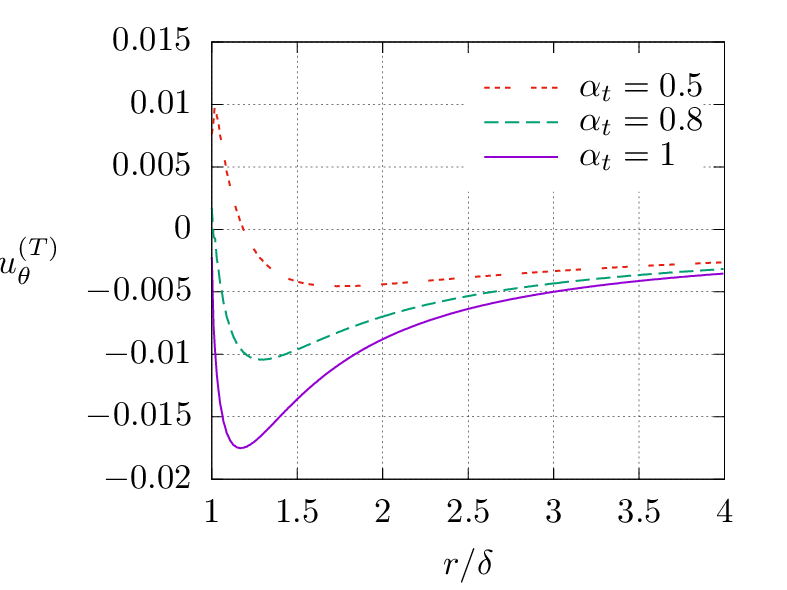}
}
\subfigure[Rarefaction parameter $\delta$=1]{
\includegraphics[scale=0.8]{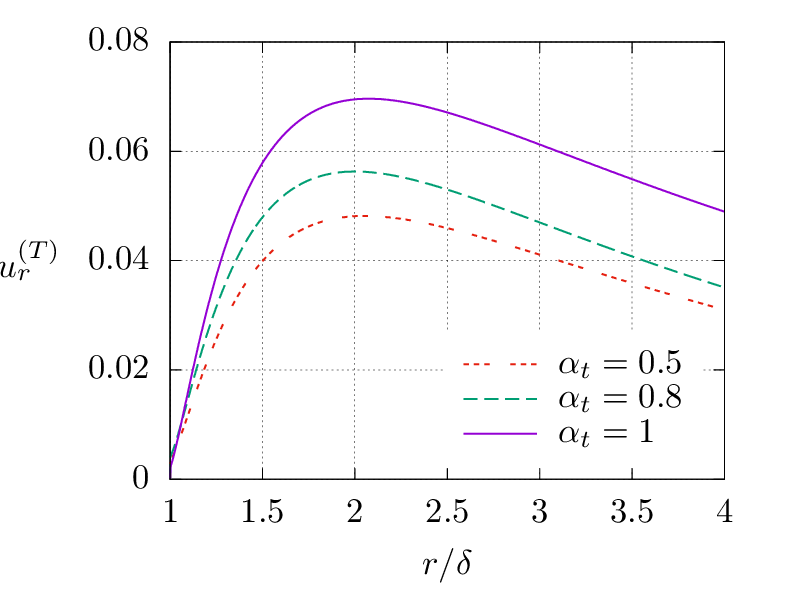}
\includegraphics[scale=0.8]{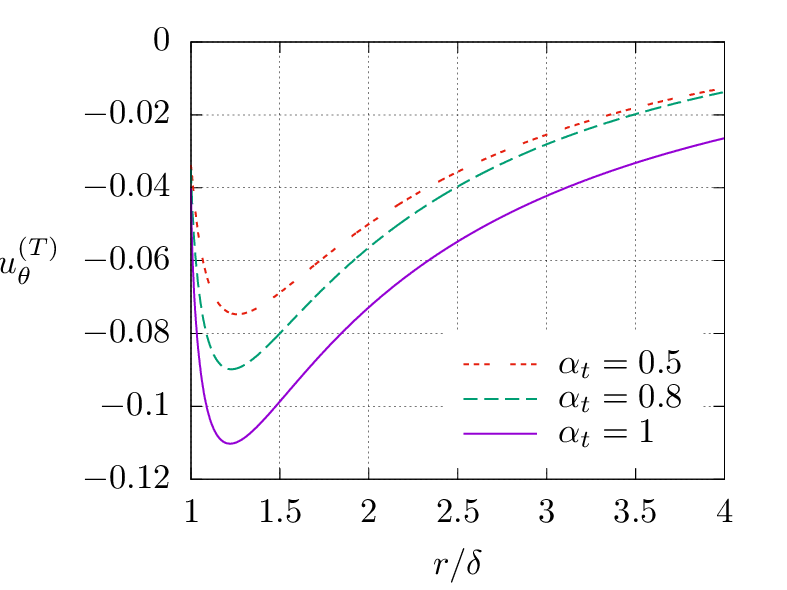}
}
\subfigure[Rarefaction parameter $\delta$=10]{
\includegraphics[scale=0.8]{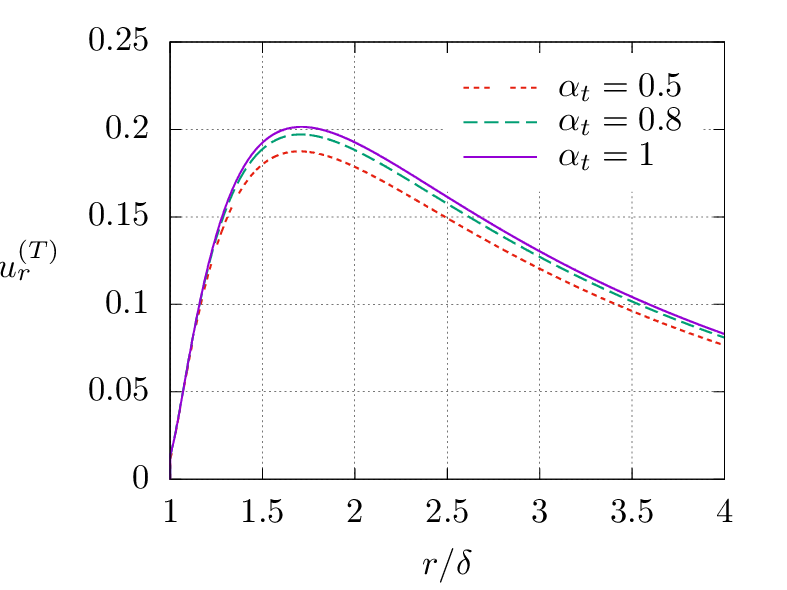}
\includegraphics[scale=0.8]{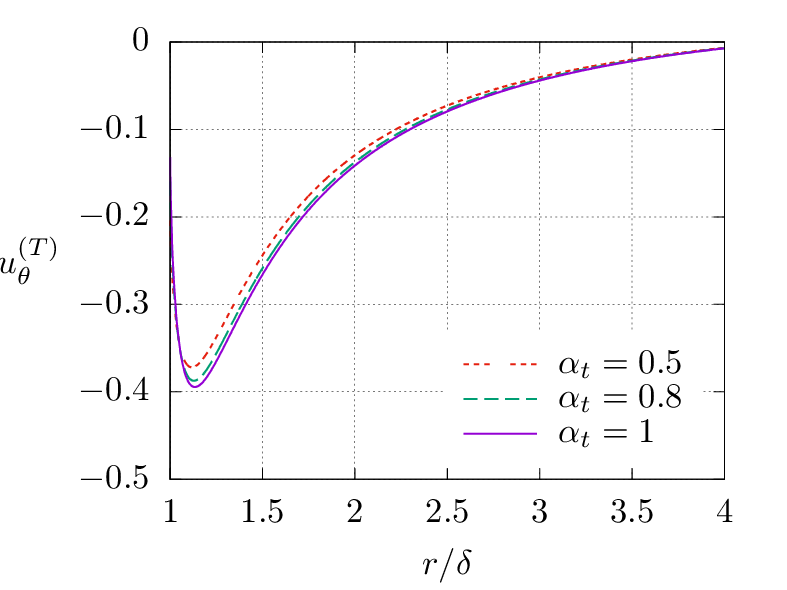}
}
\caption{Components of the bulk velocity as functions of the radial distance
from the sphere due to the
thermodynamic force $X_T$ for fixed $\alpha_n$=0.1.}
\lae{figG}
\end{figure}


\begin{figure}
\centering
\subfigure[Rarefaction parameter $\delta$=0.1]{
\includegraphics[scale=0.8]{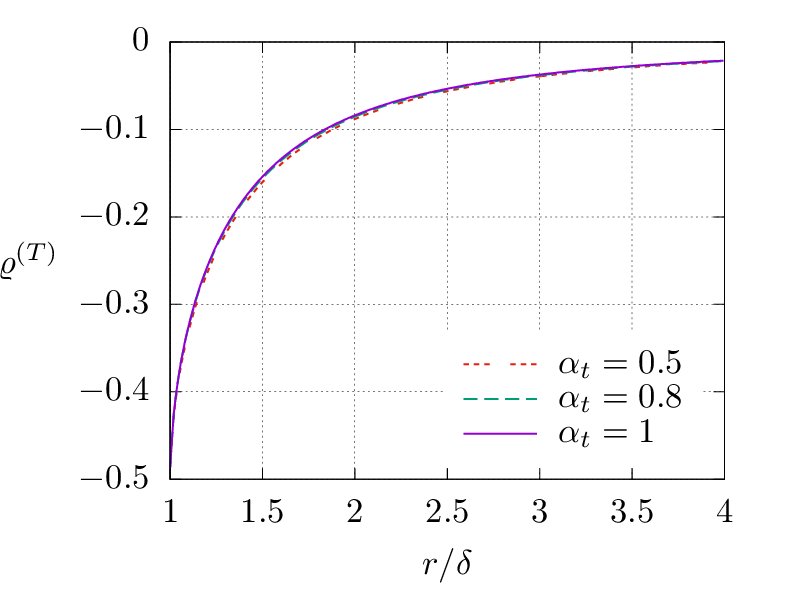}
\includegraphics[scale=0.8]{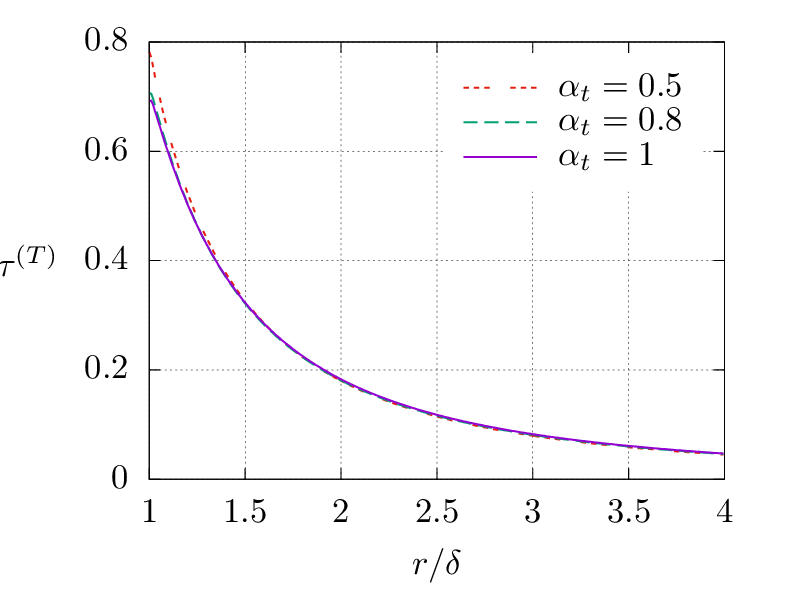}
}
\subfigure[Rarefaction parameter $\delta$=1]{
\includegraphics[scale=0.8]{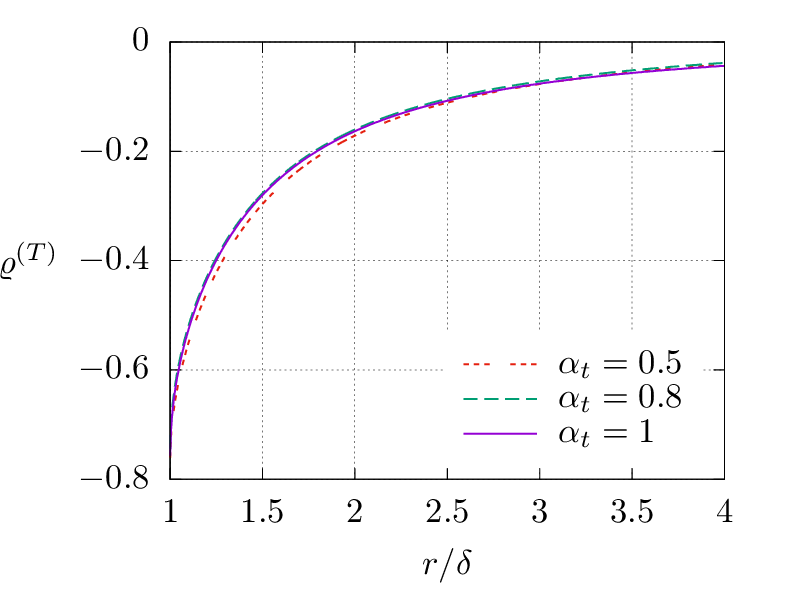}
\includegraphics[scale=0.8]{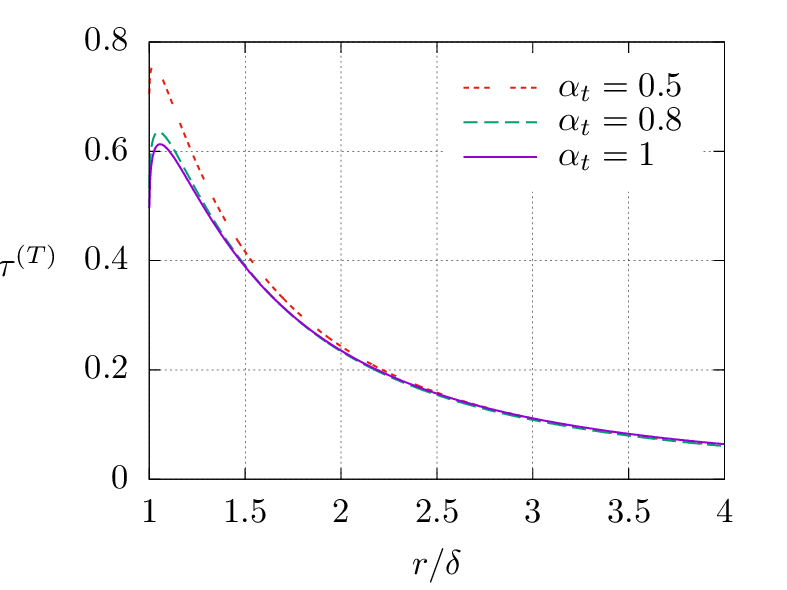}
}
\subfigure[Rarefaction parameter $\delta$=10]{
\includegraphics[scale=0.8]{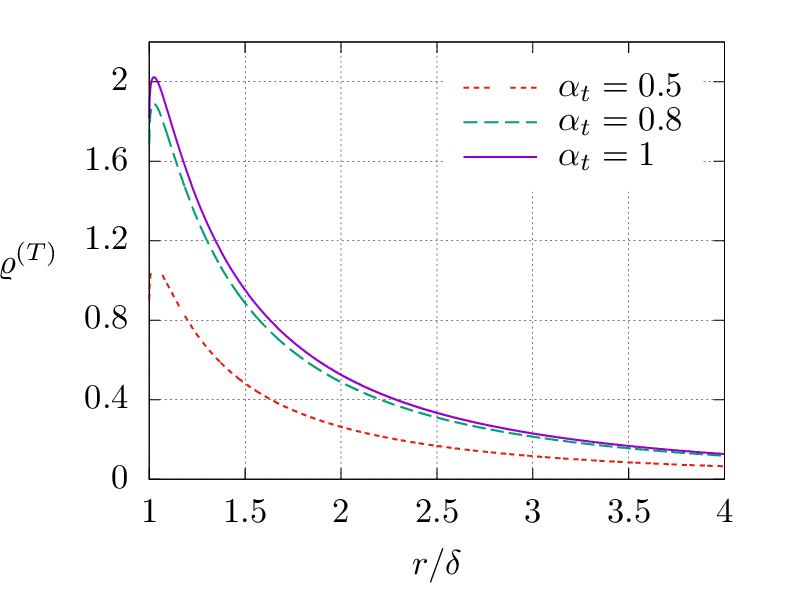}
\includegraphics[scale=0.8]{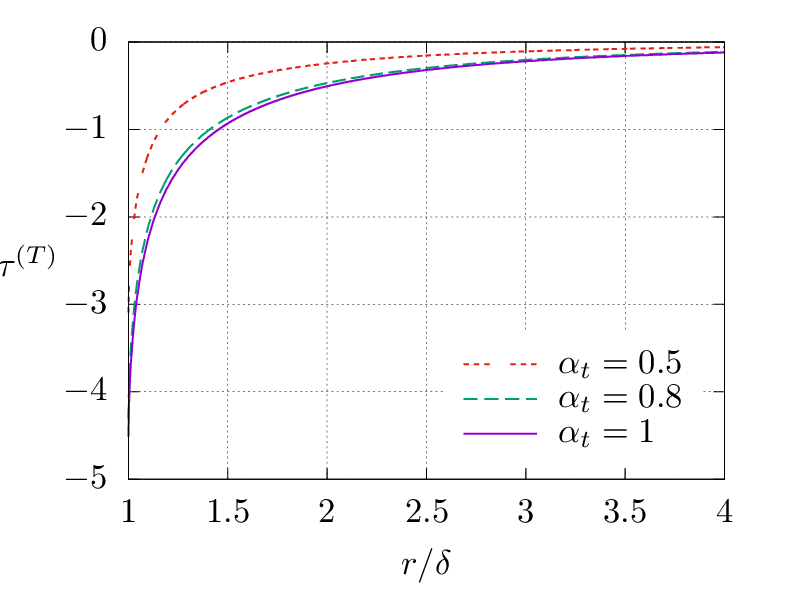}
}
\caption{Density and temperature deviations as functions of the radial
distance from the sphere due to the
thermodynamic force $X_T$ for fixed $\alpha_n$=0.1.}
\lae{figH}
\end{figure}


\begin{figure}
\centering
\subfigure[Rarefaction parameter $\delta$=0.1]{
\includegraphics[scale=0.8]{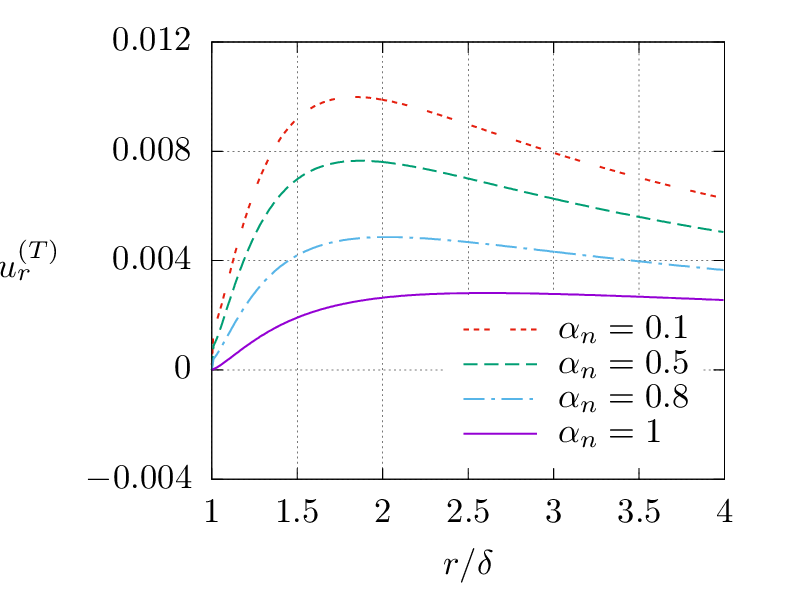}
\includegraphics[scale=0.8]{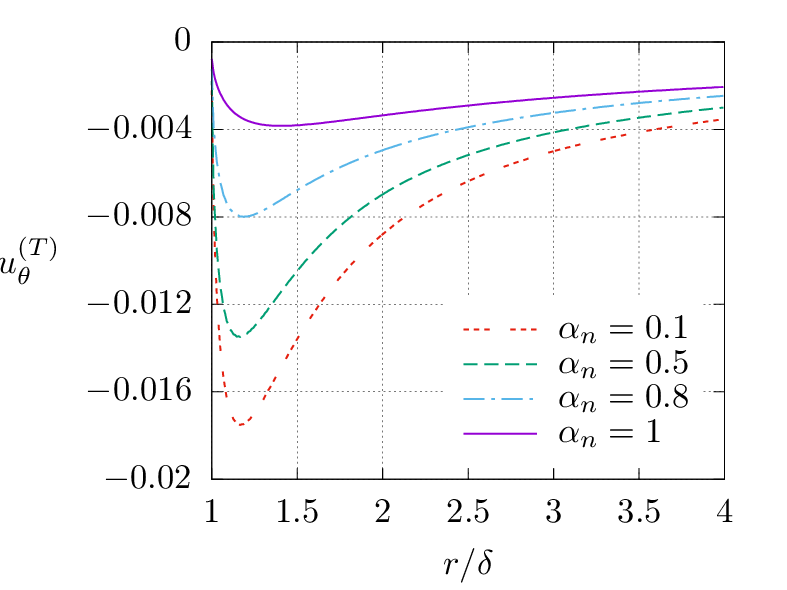}
}
\subfigure[Rarefaction parameter $\delta$=1]{
\includegraphics[scale=0.8]{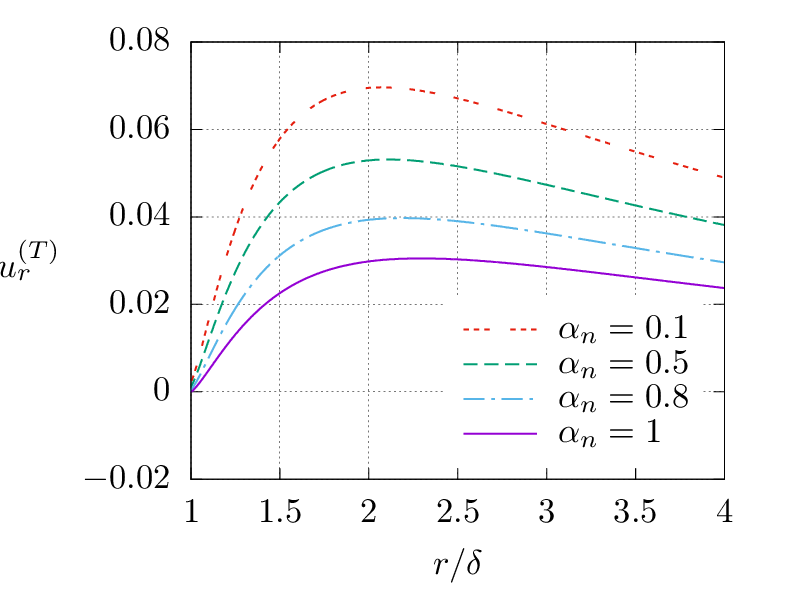}
\includegraphics[scale=0.8]{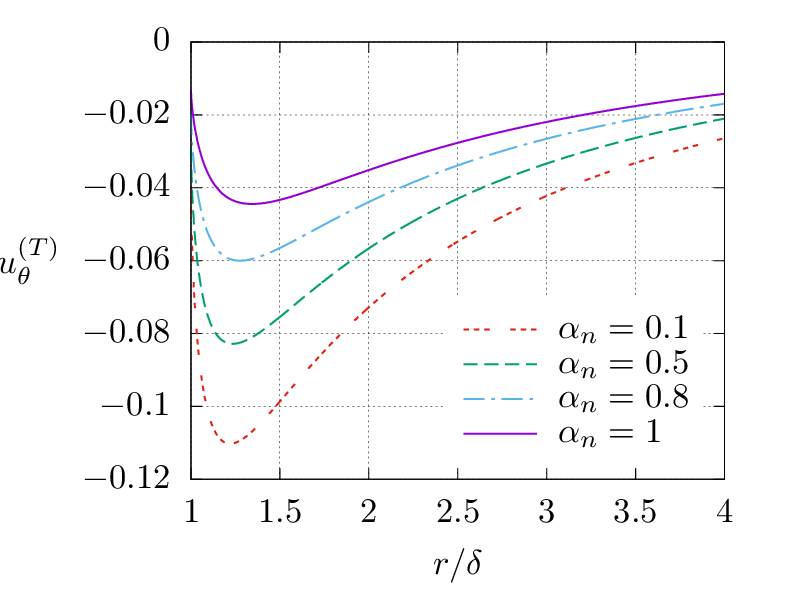}
}
\subfigure[Rarefaction parameter $\delta$=10]{
\includegraphics[scale=0.8]{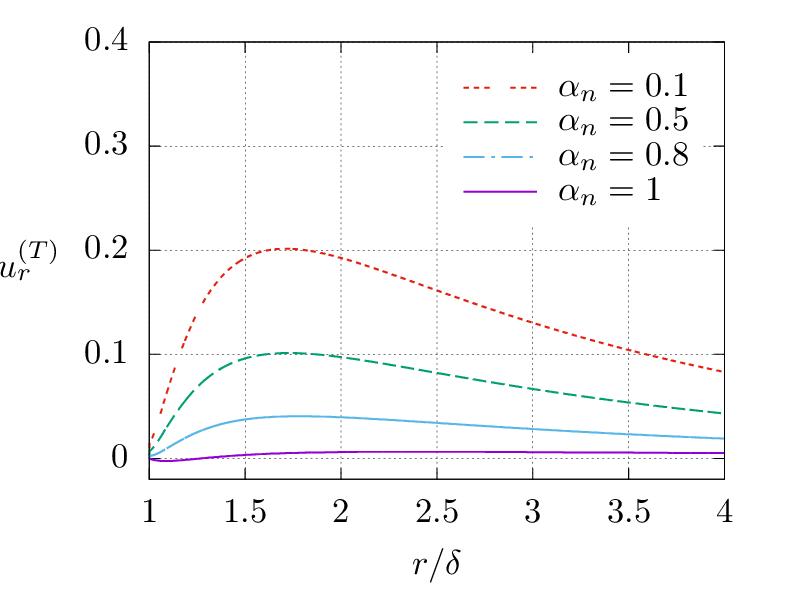}
\includegraphics[scale=0.8]{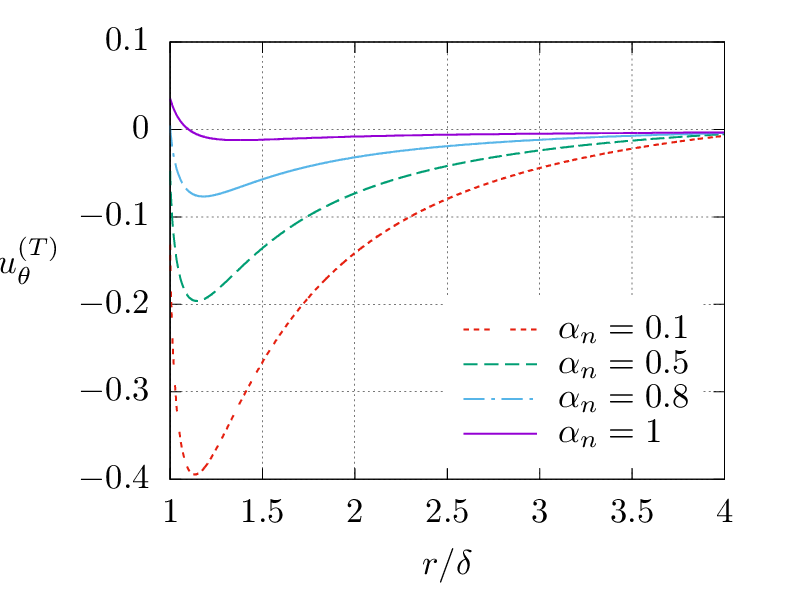}
}
\caption{Components of the bulk velocity as functions of the radial distance
from the sphere due to the
thermodynamic force $X_T$ for fixed $\alpha_t$=1.}
\lae{figI}
\end{figure}


\begin{figure}
\centering
\subfigure[Rarefaction parameter $\delta$=0.1]{
\includegraphics[scale=0.8]{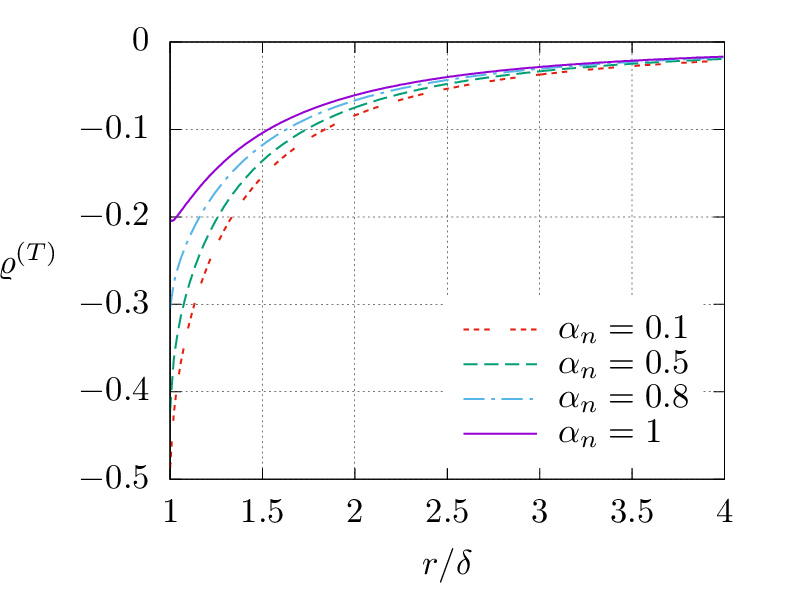}
\includegraphics[scale=0.8]{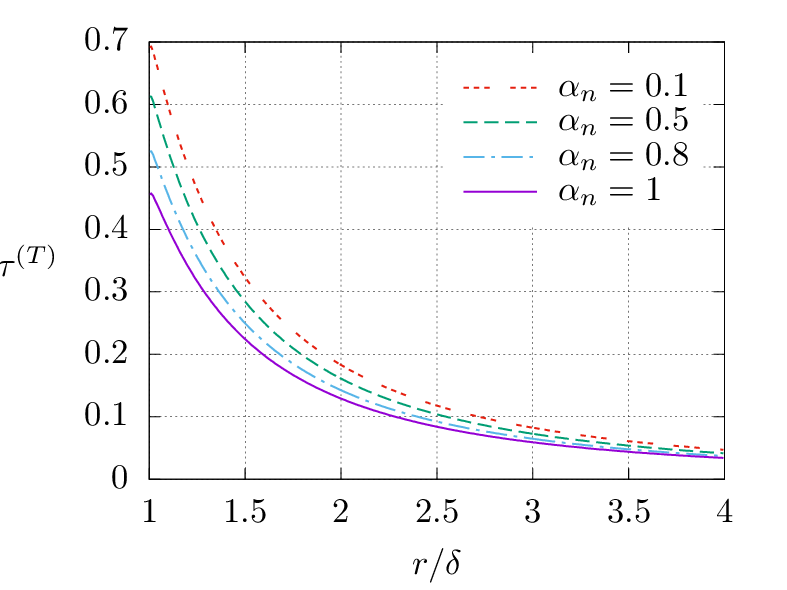}
}
\subfigure[Rarefaction parameter $\delta$=1]{
\includegraphics[scale=0.8]{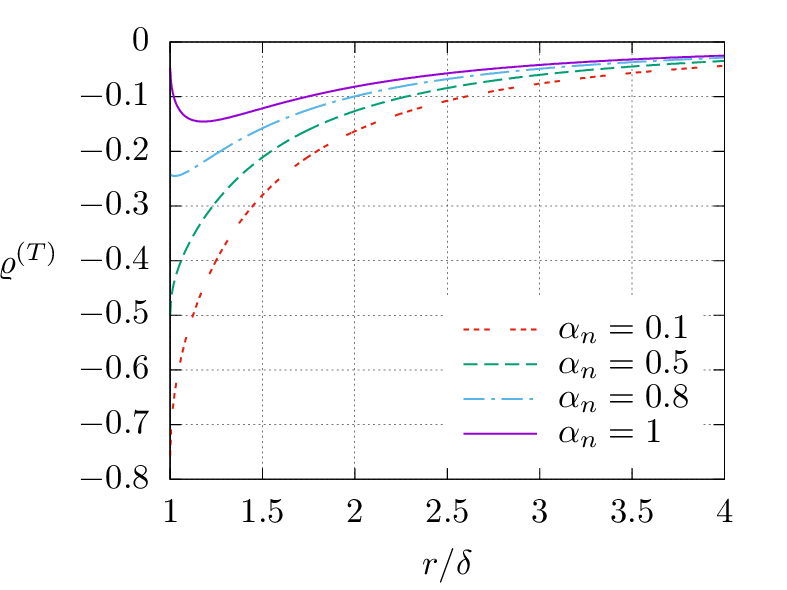}
\includegraphics[scale=0.8]{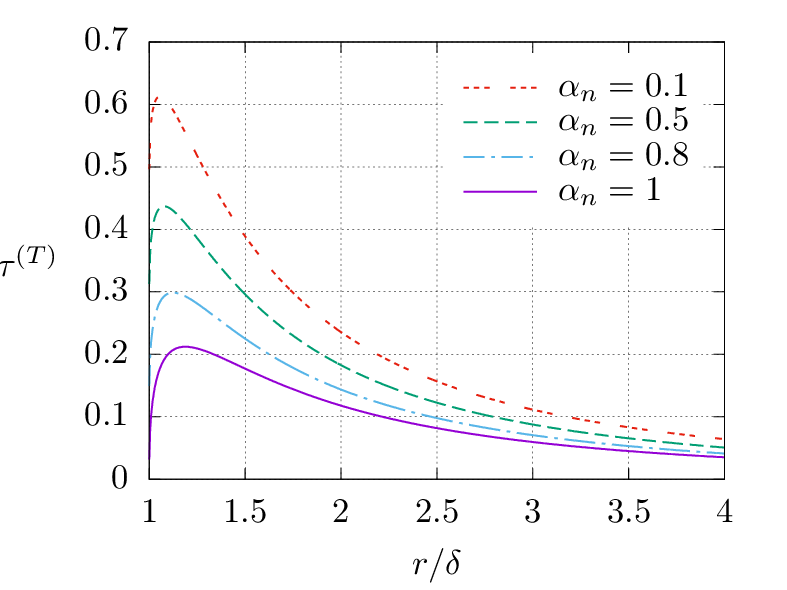}
}
\subfigure[Rarefaction parameter $\delta$=10]{
\includegraphics[scale=0.8]{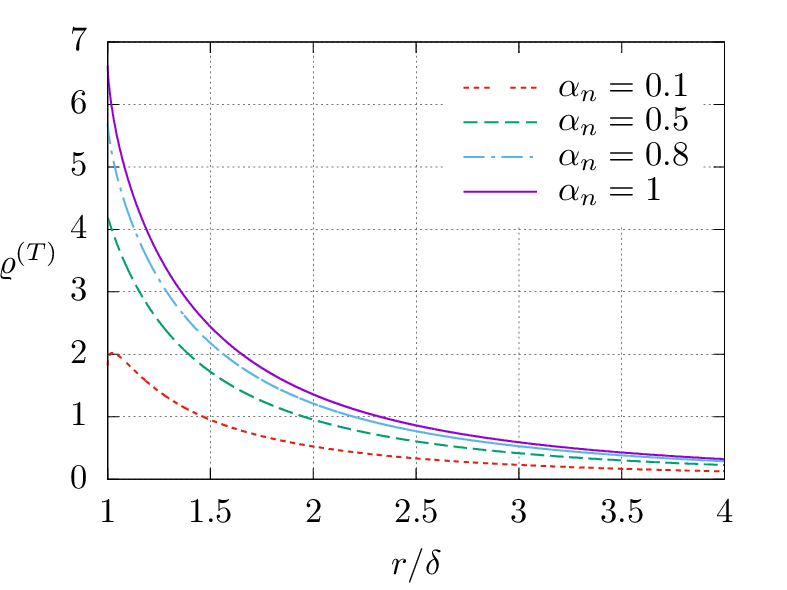}
\includegraphics[scale=0.8]{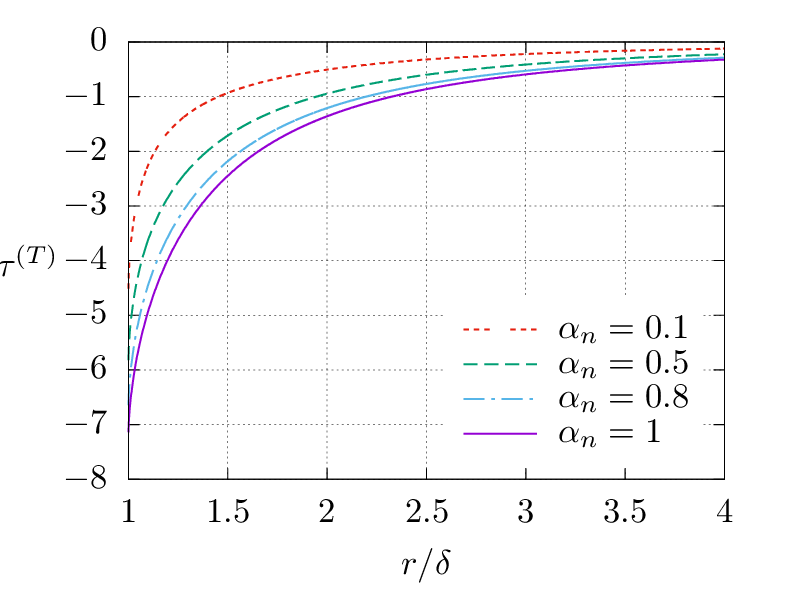}
}
\caption{Density and temperature deviations as functions of the radial
distance from the sphere due to the
thermodynamic force $X_T$ for fixed $\alpha_t$=1.}
\lae{figJ}
\end{figure}

\clearpage

\begin{figure}
\centering
\includegraphics[scale=0.8]{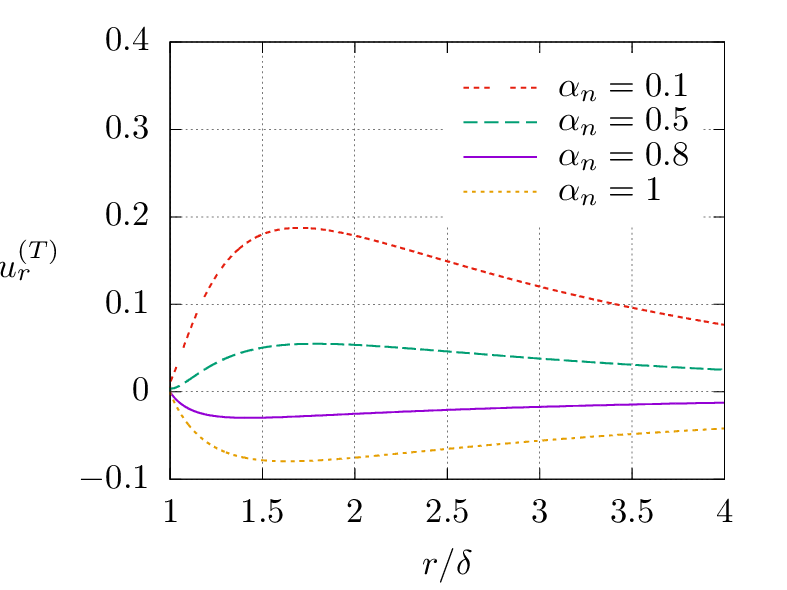}
\includegraphics[scale=0.8]{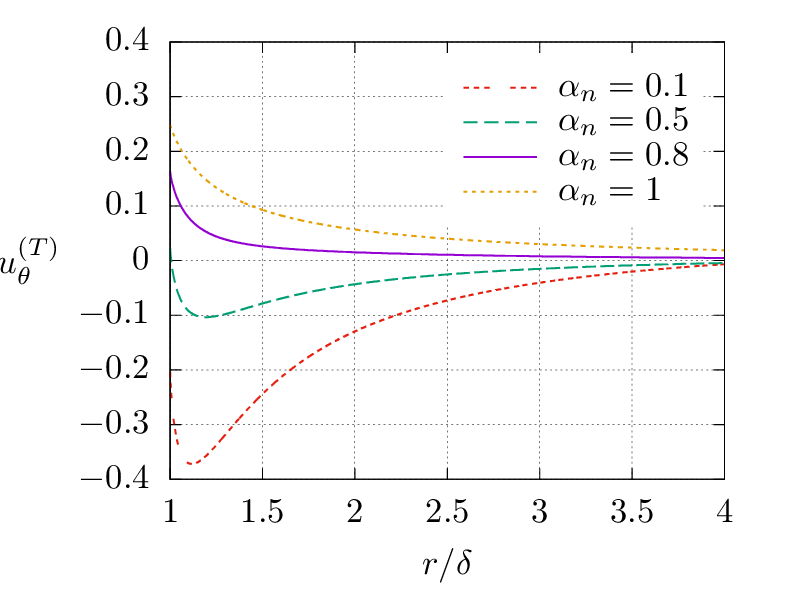}
\caption{Components of the bulk velocity as functions of the radial distance
from the sphere due to thermodynamic force $X_T$
for fixed $\alpha_t$=0.5 and $\delta$=10.}
\lae{figK}
\end{figure}

\clearpage

\begin{figure}
\centering
\includegraphics[scale=1.]{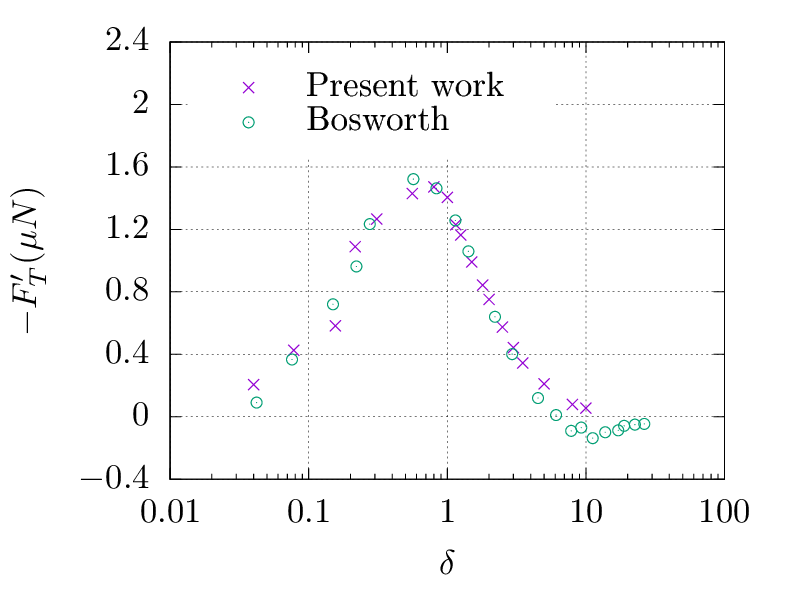}
\caption{Dimensionless thermophoretic force: comparison with experimental
data provided by \cite{Bos02} for a copper sphere in argon gas.}
\lae{figexp}
\end{figure}

\end{document}